\documentclass[11pt]{article}

\usepackage{amsmath,amssymb,amsthm,amsbsy}
\usepackage{mathrsfs} 
\usepackage{graphicx}	
\usepackage{fullpage}	
\usepackage{multirow} 
\usepackage{setspace} 

\usepackage{lineno}			
\usepackage{algorithm} 	
\usepackage{enumerate} 	
\usepackage{xfrac}			

\usepackage[comma]{natbib}		  

\usepackage[bookmarks=true, linkbordercolor={1 1 1}, citebordercolor={1 1 1}]{hyperref}

\theoremstyle{plain} \newtheorem{lemma}{Lemma}
\theoremstyle{plain} \newtheorem{proposition}{Proposition}
\theoremstyle{plain} 
\theoremstyle{definition} 
\theoremstyle{definition} 
\theoremstyle{definition} \newtheorem{assumption}{Assumption}
\theoremstyle{definition} \newtheorem{example}{Example}

\theoremstyle{plain} \newtheorem*{lemma*}{Lemma}
\theoremstyle{plain} \newtheorem*{proposition*}{Proposition}
\theoremstyle{plain} \newtheorem*{theorem*}{Theorem}
\theoremstyle{definition} \newtheorem*{corollary*}{Corollary}
\theoremstyle{definition} \newtheorem*{definition*}{Definition}
\theoremstyle{definition} \newtheorem*{assumption*}{Assumption}
\theoremstyle{definition} \newtheorem*{example*}{Example}

\newcommand{\R}{\ensuremath{\mathbb{R}}}										

\newcommand{\E}[1]{\mathbb{E}\left[#1\right]}								
\newcommand{\CE}[2]{\mathbb{E}\left[#1 \left\vert\right. #2 \right]}	
\newcommand{\Prob}[1]{\mathbb{P}\left[#1\right]}						
\newcommand{\pos}[1]{\left[#1\right]^{+}}										
\newcommand{\indicator}[1]{\textbf{1}_{\left\{#1\right\}}} 	
\newcommand{\od}[2]{\frac{d #1}{d #2}}											
\newcommand{\pd}[2]{\frac{\partial #1}{\partial #2}}				

\newcommand{\mb}[1]{\ensuremath{\mathbf{#1}}}
\newcommand{\bs}[1]{\boldsymbol #1}													
\newcommand{\mc}[1]{\ensuremath{\mathcal{#1}}}							
\newcommand{\wh}[1]{\widehat{#1}}		
\newcommand{\wt}[1]{\widetilde{#1}}		

\newcommand{\beq}[1]{\begin{equation} \label{eq:#1}}
\newcommand{\eeq}{\end{equation}}
\newcommand{\beqn}{\begin{equation*}}
\newcommand{\eeqn}{\end{equation*}}

\renewcommand{\cite}[1]{\citet{#1}}

\newcommand*\samethanks[1][\value{footnote}]{\footnotemark[#1]} 

\newcommand{\Ebig}[1]{\mathbb{E}\big[#1\big]}

\newcommand{\posb}[1]{[#1]^{+}}
\newcommand{\posbig}[1]{\big[#1\big]^{+}}

\newcommand{\eop}{\hfill \qedsymbol}

\title{\bf Auctions with a Profit Sharing Contract}

\author{
Vineet Abhishek\thanks{Department of Electrical and Computer Engineering, University of Illinois at Urbana-Champaign. \newline Contact: {\tt \small abhishe1@illinois.edu} and {\tt \small b-hajek@illinois.edu}},
~Bruce Hajek\samethanks[1]
~and Steven R. Williams\thanks{Department of Economics, University of Illinois at Urbana-Champaign. 
\newline Contact: {\tt \small swillia3@illinois.edu}}
}
\begin{document}
\date{Working paper. First draft: February 14, 2011. This draft: July 25, 2012.}
\maketitle

\begin{abstract}
We study the problem of selling a resource through an auction mechanism. The winning buyer in turn develops this resource to generate profit. Two forms of payment are considered: charging the winning buyer a one-time payment, or an initial payment plus a \textit{profit sharing contract} (PSC). We consider a symmetric interdependent values model with risk averse or risk neutral buyers and a risk neutral seller. For the second price auction and the English auction, we show that the seller's expected total revenue from the auction where he also takes a fraction of the positive profit is higher than the expected revenue from the auction with only a one-time payment. Moreover, the seller can generate an even higher expected total revenue if, in addition to taking a fraction of the positive profit, he also takes the same fraction of any loss incurred from developing the resource. Moving beyond simple PSCs, we show that the auction with a PSC from a very general class generates higher expected total revenue than the auction with only a one-time payment. Finally, we show that suitable PSCs provide higher expected total revenue than a one-time payment even when the incentives of the winning buyer to develop the resource must be addressed by the seller.

\end{abstract}

\section{Introduction} \label{sec:introduction}
Auctions are commonly used to sell resources such as licenses to use a particular wireless spectrum bandwidth or oil and gas drilling rights in a tract of land.\footnote{See \cite{Cramton97} and \cite{Milgrom04} for reviews of the FCC spectrum auctions, and \cite{Porter95} and \cite{Cramton09} for reviews of oil rights auctions.} The auction of a resource is different from the auction of an item for consumption. A resource in turn can be developed by the winning buyer to generate profit. Once a buyer develops the resource, the profit may become known to both the buyer and the seller through observable elements such as sales data, quality of the resource, market condition, etc. This raises the possibility of the seller selecting ex-ante as part of his auction a contract that specifies how the net profit will be split ex-post between the winning buyer and the seller.

We compare in this paper the following two forms of payment: (i) the seller either charges the winning buyer a one-time payment at the end of the auction stage; or (ii) the seller receives an initial payment from the winning buyer at the end of the auction stage followed by a \textit{\ profit-sharing contract} (henceforth, PSC) in which he additionally receives a prespecified share of the realized profit from the resource. This is motivated by a current example: the FCC spectrum auctions (e.g., auction 73) are of the first type, while the 3G spectrum auctions in India require that a winning buyer pay a spectrum usage charge equal to a percentage of his profit in addition to the spectrum acquisition fees.\footnote{See \cite{FCCAuction73} and \cite{India3g} for details.} We investigate whether or not there are economic reasons for the seller to prefer auctions with a PSC over auctions with only a one-time payment. The solution to this problem is nontrivial because strategic buyers adjust their bids in the auction stage in response to the payment they are required to make according to the PSC.

We address this problem using the symmetric interdependent values model of Milgrom and Weber \citeyearpar{Milgrom&Weber82}. This model includes two extremes as special cases -- the independent private value model and the pure common value model -- as well as a continuum of interdependent value models between these two extremes. The value of the resource to a buyer is not known to him before it is developed. However, each buyer has some information about the resource, known only to him. This information may be informative to other buyers as it can refine their respective estimates. The value of the resource to the winning buyer is publicly observable once it is developed by him. We work with risk averse or risk neutral buyers and a risk neutral seller; the seller of the resource is usually a large organization (e.g., the government of a country) with large cash reserves and possibly owns multiple such resources. 

Our prime focus is on two simple PSCs. First is a \textit{profit only sharing contract} (henceforth, POSC) where the seller takes a fixed fraction $\alpha \in (0,1)$ of the positive profit from the winning buyer but does not take any negative profit (loss) from him. Second is a \textit{profit and loss sharing contract} (henceforth, PLSC) where the seller takes a fraction $\alpha \in (0,1)$ of both positive and negative profit from the winning buyer. For the second price and the English auctions, we show that:

\begin{enumerate}[(i)]
\item
The seller's expected total revenue from the auction with a POSC or a PLSC is nondecreasing in the share fraction $\alpha \in (0,1)$, and in particular, is higher than the expected revenue from the auction with only a one-time payment (i.e., $\alpha = 0$). Hence, there are economic reasons to prefer a POSC or a PLSC over only a one-time payment.

\item
For the same share fraction $\alpha \in (0,1)$, the auction with a PLSC generates higher expected total revenue than the auction with a POSC. We show that the revenue superiority of the PLSC for a given~$\alpha$ over the corresponding POSC primarily stems from the joint effect of positive dependence between the values and the signals that is assumed in this paper, and the nature of PLSC and POSC; the effect of risk aversion is aligned with these factors. This revenue superiority is shown to hold strictly in an example in Section \ref{sec:rn-sp} with risk neutral buyers. 

\item
We leverage the intuition gained from the analysis of the POSC and PLSC to move beyond simple PSCs and show that the auction with a PSC from a very general class generates higher expected total revenue than the auction with only a one-time payment. Moreover, the PLSC with the share fraction $\alpha$ is revenue optimal over the general class of PSCs for which the seller's marginal share is bounded by $\alpha$. The seller can therefore accomplish all of his goals in revenue enhancement using linear sharing contracts.
\end{enumerate}

An obvious question is: ``If the value of the resource ultimately becomes known to the seller, then why not simply tax ex-post the entire profit from the resource, subject only to participation constraints on the buyers?'' This scenario in fact results by letting the share fraction $\alpha$ in the POCS or PLSC approach one.\footnote{The case of $\alpha =1$ is mathematically degenerate in our model. Since the seller takes the entire profit ex-post from the winning buyer, a buyer will either decline to participate or place any bid that gives him zero profit. The dependence of a buyer's bid upon his signal in our analysis of the case of $\alpha <1$ therefore does not apply for $\alpha =1$.} There are several reasons why it is worthwhile to consider a share fraction~$\alpha$ less than one and other nontrivial PSCs between the seller and the winning buyer, some of which are enumerated in Section \ref{sec:discussion}. A primary reason is that development of the resource requires effort or expertise from the winning buyer; otherwise, the seller might not need to sell the resource. Section \ref{sec:pa} models the relationship between the winning buyer and the seller using a principal-agent model. We show through an example that the revenue maximizing share fraction~$\alpha$ in either a POSC or a PLSC can be strictly less than one because of the necessity of providing proper incentives to the winning buyer to develop the resource. However, a one-time payment is never revenue optimal: we show that in our principal-agent model, the auction with a PLSC and a suitably small share fraction $\alpha$ generates higher expected total revenue than the auction with only a one-time payment.\footnote{As explained in Section \ref{sec:pa}, it is difficult to analyze more general PSCs or even the POSC in a principal-agent relationship. A numerical example in Section \ref{sec:pa} suggests, however, that: (i) as with the PLSC, a POSC with sufficiently small share fraction $\alpha$ produces more expected revenue for the seller than a one-time payment; (ii) for fixed $\alpha$, a POSC causes the winning buyer to choose a larger effort than the corresponding PLSC and may thereby produce a larger expected revenue for the seller. The generality of these two observations remains to be explored.} The advantages of the PSCs thus carry over to the case in which seller must provide the winning buyer with incentives to develop the resource.

\textbf{Related work:} There are a number of approaches in the literature to increase the seller's revenue, either by exploiting the informational structure or by making payments contingent on the ex-post realization of the values. The distinguishing aspects of our work are: allowing for risk aversion among buyers; a substantially more general informational environment;\footnote{While most of the revenue ranking results in auction theory are through applications of the \textit{linkage principle} (see \cite{Krishna2002}) and assume various smoothness conditions, our proofs rely entirely on the properties of concave functions and stochastic ordering with no or only mild assumptions of smoothness.} consideration of very general transfers ex-post; analysis for the second price and the English auctions; and consideration of the incentives of the winning buyer to develop the resource. 

There are two sets of papers especially relevant to our work. The first set of papers use a mechanism design approach to maximize revenue for the seller. For the cases of statistically dependent private values, a common value model in which the types of the buyers are independent conditional on the common value, and interdependent values with statistically dependent types, the seller can extract the full surplus from the buyers using one-time payments.\footnote{See \cite{Cremer&McLean88,Cremer&McLean85}, \cite{McAfee&McMillan&Reny89}, and \cite{McAfee&Reny92} for details.} The Wilson critique of mechanism design (\cite{Wilson87}) is useful for distinguishing our approach from the papers on full surplus extraction. While these papers define transfers between the seller and the buyers in terms of the probability distribution of their types, our approach instead uses two common auctions that are defined without reference to the buyers' beliefs.\footnote{Except in the special case of independent private values, however, the two auctions that we consider require that the buyers have common knowledge of their beliefs in formulating their bids. The work on full surplus extraction made honest reporting into a Bayesian-Nash equilibrium; while making an honest report does not require that common knowledge of beliefs among buyers, this common knowledge is required if a buyer is to verify that honest reporting is incentive compatible.} Along with a fractional sharing contract, these auctions are the sort of simple and familiar trading procedures that Wilson advocates. 

\cite{Mezzetti07} considers a model in which the value of the item to any risk neutral buyer is a function of his type and the types of the other buyers; a buyer who receives the item realizes its value ex-post. \citeauthor{Mezzetti07} proposes a two-stage reporting procedure in which all buyers first report their types followed by a subsequent report by the winning buyer of the item's realized value. All losing buyers are compelled to pay a large fine ex-post if the reported value of the winning buyer is inconsistent with the earlier reports by the losing buyers. This insures honest reporting by all buyers and allows full surplus extraction. Practically, however, one must question the willingness of buyers to participate in a mechanism in which they may be required to pay a large fine ex-post depending upon: (i) the reports of the other buyers in the first stage; (ii) the ex-post report of the winning buyer who, while having no incentive to lie, also has no incentive to be honest.

The second set of papers use ex-post transfers to enhance the revenue of the seller. \cite{Hansen85} constructs examples of an oil tract sale and the sale of a firm to other firms. It is shown in each example that the seller benefits by retaining a share of the profit from the future enterprise. The examples are restricted to the cases of independent private values with risk neutral buyers. \cite{Riley88} shows in a symmetric interdependent values model that if the value of a resource is observed ex-post, even if imperfectly, then the seller can increase his expected revenue in a first price auction by either requiring a winning buyer to pay a predetermined royalty or by having the buyer establish a royalty rate through bidding. Along with using an auction mechanism different from what we use, \citeauthor{Riley88}'s analysis is restricted to risk-neutral buyers and linear ex-post transfers.

Ex-post information is also used by \cite{MaCfee&McMillan86} and \cite{Laffont&Tirole87} to analyze procurement auctions in which the winning bidder privately exerts effort to reduce his costs. This raises the principal-agent issue that is mentioned above. These papers demonstrate that the auctioneer can profit from contracting a cost-sharing arrangement with the winning bidder. Both papers, however, are limited to the case of independent private values, which is ill-suited to modeling the sale of resources such as spectrum bandwidth or mineral rights. Moreover, each paper assumes specific utility functions for the bidder.\footnote{\cite{MaCfee&McMillan86} assumes that each bidder either has the same constant absolute risk aversion (CARA) function as his utility of money or each bidder is risk neutral, while \cite{Laffont&Tirole87} assumes risk neutrality.} \citeauthor{MaCfee&McMillan86}'s cost sharing contract is closely related to the PLSC, though they consider the first price auction. Our analysis of the PLSC in a principal-agent relationship is for the second price auction and the English auction. 

\cite{DeMarzo2005} considers the first and the second price auctions in which bids are selected from a completely ordered family of securities whose ultimate values are tied to the resource being auctioned. \citeauthor{DeMarzo2005} define a partial ordering based on the notion of \textit{steepness} and show that the steeper family of securities provides higher expected revenue to the seller. The two stage payment rules that we consider can be viewed as securities and our revenue ranking of different PSCs is consistent with their idea of steepness. Our paper is more general in allowing for risk aversion among buyers and in its substantially less restrictive informational assumptions. The generality of our informational model is significant for more than mathematical reasons: \cite{Abhishek-etal2011b} shows that the revenue ranking of \cite{DeMarzo2005} does not necessarily hold if its informational environment is relaxed. Thus, while our results are consistent with theirs, they do not follow from their results. We also complement their work by analyzing the English auction.

\textbf{Outline of this paper:} The rest of this paper is organized as follows. Section \ref{sec:model} outlines our model, notation, and definitions. Section \ref{sec:second-price} analyzes the second price auction with a POSC and with a PLSC, obtains an equilibrium strategy of the buyers, and establishes the revenue consequences of these two PSCs. Section \ref{sec:english} extends the results of Section \ref{sec:second-price} to the English auction. Section~\ref{sec:discussion} provides comments and extensions of our model. Section \ref{sec:general-psf} studies the revenue consequences of general PSCs. Section \ref{sec:pa} analyzes auctions with PSCs in a principal-agent relationship. We conclude in Section \ref{sec:conclusions}.

\section{Model and Notation} \label{sec:model}
Consider $N$ buyers competing for a resource that a seller wants to sell. The value $x_n$ of the resource to a buyer $n$ is a realization of a random variable $X_n$, unknown to him. This is the profit to buyer~$n$ from developing the resource in the absence of any payments to the seller, but after taking into account the variable costs. The $X_n$'s can take negative values, with the interpretation as loss incurred from developing the resource. Buyer $n$ privately observes a signal $y_n$ through a realization of a random variable $Y_n$ that is correlated with $(X_1, X_2, \ldots, X_N)$. The joint cumulative distribution function (CDF) of the random variables $X_n$'s and $Y_n$'s is common knowledge.

Let $\mb{x} \triangleq (x_1, x_2, \ldots, x_N)$ denote a vector of values; denote the random vector $(X_1, X_2, \ldots, X_N)$ by \mb{X}. A vector of signals $\mb{y}$ and the random vector $\mb{Y}$ are defined similarly.  We use the standard game theoretic notation of $\mb{x}_{-n} \triangleq (x_1, \ldots, x_{n-1}, x_{n+1}, \ldots, x_N)$. Similar interpretations are used for $\mb{X}_{-n}$, $\mb{y}_{-n}$, and $\mb{Y}_{-n}$. Let $F_{\mb{X},\mb{Y}}(\mb{x},\mb{y})$ denote the joint CDF of $(\mb{X}, \mb{Y})$. It is assumed to have the following symmetry property:

\begin{assumption} \label{assumption:symmetry}
The joint CDF of $(X_n, Y_n, \mb{Y}_{-n})$, denoted by $F_{X_n, Y_n, \mb{Y}_{-n}}(x_n, y_n, \mb{y}_{-n})$, is identical for each $n$ and is symmetric in the last $N-1$ components $\mb{y}_{-n}$.
\end{assumption}

Assumption \ref{assumption:symmetry} allows for a special dependence between the value of the resource to a buyer and his own signal, while the identities of other buyers are irrelevant to him. The model reduces to the independent private values model if $(X_n, Y_n)$ is independent of $(\mb{X}_{-n},\mb{Y}_{-n})$ for all~$n$, 
to the pure common value model if $X_1 = X_2 = \ldots = X_N$, and includes a continuum of interdependent value models between these two extremes. Because of Assumption~\ref{assumption:symmetry}, the subsequent assumptions and the analysis in the paper is given from buyer~$1$'s viewpoint.

The set of possible values that each random variable $Y_n$ can take is assumed to be an interval $\mc{I}_{Y} \subset \R$. Assume that the joint probability density function (pdf) of the random vector $\mb{Y}$, denoted by $f_{\mb{Y}}(\mb{y})$, exists and is positive for all $\mb{y} \in \mc{I}_{Y}^N$. Notice that $f_{\mb{Y}}(\mb{y})$ is symmetric in its $N$ arguments. The random variables $(X_n, Y_n, \mb{Y}_{-n})$ are not required to have a joint pdf. In particular, this allows $X_n$ to take discrete values or to be a deterministic function of $(Y_n, \mb{Y}_{-n})$.

Let larger numerical values of the signals correspond to more favorable estimates of the value of the resource. Mathematically, we assume the following form of positive dependence between the value of the resource and the buyers' signals: 

\begin{assumption} \label{assumption:pos-dep}
For any increasing function $h:\R \mapsto \R$, $a_n$'s, and $b_n$'s such that $a_n, b_n \in \mc{I}_{Y}$ and $a_n \leq b_n$ for all $n$, the conditional expectation $\E{h(X_1) | Y_1 \in [a_1,b_1], Y_2 \in [a_2,b_2], \ldots, Y_N \in [a_N,b_N]}$ is increasing in $a_1$ and $b_1$, and nondecreasing in $\mb{a}_{-1}$ and $\mb{b}_{-1}$ whenever it exists.\footnote{Throughout this paper, ``increasing'' means ``strictly increasing'' and ``decreasing'' means ``strictly decreasing''. While the results of the paper extend to the case in which the conditional expectation in Assumption \ref{assumption:pos-dep} is nondecreasing in the $a_{n}$'s and $b_{n}$'s, the analysis is more complicated.}
\end{assumption}

Assumption \ref{assumption:pos-dep} trivially implies that the random variable $X_1$ conditioned on $\mb{Y} = \mb{y}$ is increasing in~$y_1$ and nondecreasing in~$\mb{y}_{-1}$ in the sense of first order stochastic dominance (henceforth, FOSD). Assumption \ref{assumption:pos-dep} provides a common sufficient condition for the existence of a pure strategy equilibrium in both the second price and the English auctions; it actually can be relaxed into two weaker conditions based on FOSD, one for the analysis of the second price auction (Lemma \ref{lemma:fosd-sp}) and one for the English auction (Lemma \ref{lemma:fosd-eng}).\footnote{Assumption \ref{assumption:pos-dep} is weaker than affiliation, which is commonly assumed in the auction literature.}

An auction with a PSC has two stages: 
\begin{enumerate}[(i)]
\item
\textbf{The auction stage}:
The first stage is an auction to decide who should get the resource and to determine the initial payment to the seller. We focus on the second price sealed bid auction (henceforth, just the \textit{second price auction}) and the English auction.\footnote{There are many variants of the English auction. We work with the one used in \cite{Milgrom&Weber82}.} In a second price auction, the buyer with the highest bid wins and pays the amount equal to the second highest bid. An English auction is an ascending price auction with a continuously increasing price. At each price level, a buyer decides whether to drop out or not. The price level and the number of active buyers are publicly known at any time. The auction ends when the second to last buyer drops out and the winner pays the price at which this happens.

\item
\textbf{The profit sharing stage}:
The second stage is a PSC. We use the term \textit{preliminary profit} to denote the net obtained after subtracting the auction stage payment made by the winning buyer from the value of the resource. The seller takes a fixed fraction of the preliminary profit (or loss) from the winning buyer.\footnote{For simplicity, we only consider at this point single period profit realization with no discounting. Profit realization over multiple time periods with discounting is discussed in Section \ref{sec:discussion}.} We assume that the preliminary profit is observed by both the seller and the winning buyer.

A POSC and a PLSC are each specified by a share fraction $\alpha \in [0,1)$, known to the buyers in the auction stage. If the winning buyer makes a payment~$b$ at the end of the auction stage and the value of the resource is revealed to be $x$, then the payment he makes in the second stage in the POSC is $\alpha \pos{x - b}$ (here, $\pos{a} \triangleq \max\{0,a\}$), while the payment in the PLSC is $\alpha (x - b)$. Notice that $\alpha = 0$ corresponds to having no profit sharing stage, i.e., the winning buyer makes only a one-time payment at the end of the auction stage. 

\end{enumerate}

The buyers are assumed to be risk averse or risk neutral. Each buyer has the same von Neumann-Morgenstern utility of money, denoted by $u: \R \rightarrow \R$, which is concave (possibly linear, as in the case of risk neutrality), increasing, and normalized so that $u(0) = 0$. Henceforth, we use the term \textit{weakly risk averse} to refer to risk averse or risk neutral behavior. The utility function~$u$ is over the total profit from the two stages. The seller is assumed to be risk neutral.

In what follows, we assume that all expectations and conditional expectations of interest exist and are finite. Moreover, conditioned on any signal vector $\mb{y} \in \mc{I}_{Y}^N$, the expected utility of a buyer from developing the resource without any payments is assumed to be positive; i.e., $\E{u(X_1)|\mb{Y} = \mb{y}} > 0$. Thus, the buyers who are competing for the resource expect to make a positive profit from developing it and therefore willing to participate.

Finally, define the random variables $Z_1, Z_2, \ldots, Z_{N-1}$ to be the largest, second largest, $\ldots$, smallest among $Y_2, Y_3, \ldots, Y_N$. Let $\mb{Z} \triangleq (Z_1, Z_2, \ldots, Z_{N-1})$ and denote a realization of $\mb{Z}$ by $\mb{z}$. Henceforth, in any further usage, $x_1$, $y_m$, and $z_n$ are always in the support of random variables $X_1$, $Y_m$, and $Z_n$, respectively, for $1 \leq m \leq N$ and $1 \leq n \leq N-1$; $\mb{y}$ and $\mb{z}$ are always in the support of the random vectors $\mb{Y}$ and $\mb{Z}$, respectively.

\section{The Second Price Auction with a Profit Sharing Contract} \label{sec:second-price}
This section characterizes the equilibrium bidding strategies in the second price auction with a POSC and a PLSC, and then evaluates the revenue consequences for the seller. We look for a symmetric equilibrium.

The following lemma is a consequence of Assumption \ref{assumption:pos-dep} and is used extensively in this section.
\begin{lemma} \label{lemma:fosd-sp}
$\E{h(X_1)|Y_1 = y_1, Z_1 = z_1}$ is increasing in $y_1$ and nondecreasing in $z_1$ for any increasing function $h:\R \mapsto \R$ for which the expectation exists.
\end{lemma}
\begin{proof}
From the definition of $Z_1$, $Z_1 = z_1$ if and only if at least one of the $Y_n$'s, $2 \leq n \leq N$, is equal to~$z_1$. Since $f_{\mb{Y}}(\mb{y})$ exists and is positive everywhere, conditioned on $Z_1 = z_1$, the probability that two or more $Y_n$'s are equal to $z_1$ is zero. Hence, the event $Z_1 = z_1$ can be thought of as the union of $N-1$ disjoint events, one for each $n$, $2\leq n \leq N$, such that $Y_n = z_1$. By Assumption \ref{assumption:symmetry}, conditioned on $Z_1 = z_1$, each of these events occur with equal probability. Therefore,
\begin{align*}
\E{h(X_1) | Y_1 = y_1, Z_1 = z_1}
& = \E{h(X_1) | Y_1 = y_1, Y_2 = z_1, Y_3 < z_1, Y_4 < z_1, \ldots, Y_N < z_1},\\
& = \E{h(X_1) | Y_1 = y_1, Y_2 = z_1, Y_3 \leq z_1, Y_4 \leq z_1, \ldots, Y_N \leq z_1}.
\end{align*}
The result then immediately follows from Assumption \ref{assumption:pos-dep}.
\end{proof}

\subsection{Profit only sharing contract (POSC)} \label{sec:posc-sp}
In a POSC, the winning buyer pays a fraction of any preliminary positive profit to the seller. However, the seller does not bear any share of a loss. We start by defining a function $s: [0,1) \times \mc{I}_{Y}^2 \mapsto \R $ that will be used to characterize the bidding strategies of the buyers:
\beq{s-posc-sp}
s(\alpha, y_1, z_1) \triangleq \left\{b: \E{u(X_1 - b - \alpha\pos{X_1-b}) | Y_1 = y_1, Z_1 = z_1} = 0 \right\}.
\eeq
Since $u(X_1 - b - \alpha\pos{X_1-b})$ is decreasing and continuous in~$b$, there is a unique $b$ that makes the expectation in \eqref{eq:s-posc-sp} equal to zero. The function $s(\alpha, y_1, z_1)$ is therefore well-defined. It can be interpreted as follows. If buyer $1$ makes a payment $b$ in the auction stage, his payment in the POSC stage will be $\alpha\pos{X_1-b}$ and his total profit will be $X_1 - b - \alpha\pos{X_1-b}$. Thus, $s(\alpha, y_1, z_1)$ is the payment in the auction stage at which the overall expected utility of buyer~$1$ is zero, conditioned on his signal being $y_1$, and the highest signal of other buyers being $z_1$.

The next lemma characterizes some important properties of $s(\alpha, y_1, z_1)$.
\begin{lemma} \label{lemma:s-posc-sp-properties}
The function $s(\alpha, y_1, z_1)$ is increasing in $y_1$, nondecreasing in $z_1$, decreasing in~$\alpha$, and positive for small values of $\alpha$. Moreover, for all $y_1$, $z_1$, and $\alpha \in [0,1)$,
\beq{s-posc-sp-ineq}
s(\alpha, y_1, z_1) \leq \E{X_1 | Y_1 = y_1, Z_1 = z_1},
\eeq
and the inequality is strict everywhere unless $\alpha = 0$ and $u$ is linear.
\end{lemma}
\begin{proof}
Since $X_1 - b - \alpha\pos{X_1-b}$ is increasing in $X_1$ and $u$ is increasing, Lemma \ref{lemma:fosd-sp} implies that $\E{u(X_1 - b - \alpha\pos{X_1-b}) | Y_1 = y_1, Z_1 = z_1}$ is increasing in $y_1$ and nondecreasing in $z_1$. It immediately follows from \eqref{eq:s-posc-sp} that $s(\alpha, y_1, z_1)$ is increasing in $y_1$ and nondecreasing in $z_1$. Again, since $X_1 - b - \alpha\pos{X_1-b}$ is decreasing in $\alpha$ for $X_1 > b$ and nonincreasing in $\alpha$ for $X_1 \leq b$, \eqref{eq:s-posc-sp} implies that $s(\alpha, y_1, z_1)$ is decreasing in $\alpha$. 

We assumed in Section \ref{sec:model} that $\E{u(X_1)| \mb{Y} = \mb{y}} > 0$, and so
\beqn 
\E{u(X_1)|Y_1 = y_1, Z_1 = z_1} = \E{\E{u(X_1)| \mb{Y} = \mb{y}} |Y_1 = y_1, Z_1 = z_1} > 0.
\eeqn 
It follows from \eqref{eq:s-posc-sp} that $s(0, y_1, z_1) > 0$. It is easy to see that $s$ is continuous with respect to $\alpha$. Hence, $s(\alpha, y_1, z_1)$ is positive for small values of $\alpha$.

Finally, \eqref{eq:s-posc-sp} and an application of Jensen's inequality gives
\begin{align*}
& \E{u\left(X_1 - s(\alpha, y_1, z_1) - \alpha \pos{ X_1- s(\alpha, y_1, z_1) } \right) | Y_1 = y_1, Z_1 = z_1} = 0, \\
\Rightarrow & u\left(\E{X_1 - s(\alpha, y_1, z_1) - \alpha \pos{ X_1- s(\alpha, y_1, z_1) } | Y_1 = y_1, Z_1 = z_1}\right) \geq 0, \\
\Rightarrow & \E{X_1 - s(\alpha, y_1, z_1) - \alpha \pos{ X_1- s(\alpha, y_1, z_1) } | Y_1 = y_1, Z_1 = z_1} \geq 0, \\
\Rightarrow & s(\alpha, y_1, z_1) \leq \E{X_1 | Y_1 = y_1, Z_1 = z_1} - \alpha \E{\pos{X_1-s(\alpha, y_1, z_1)} | Y_1 = y_1, Z_1 = z_1}, \\
\Rightarrow & s(\alpha, y_1, z_1) \leq \E{X_1 | Y_1 = y_1, Z_1 = z_1},
\end{align*}
where the first inequality is strict if $u$ is strictly concave and the last inequality is strict if $a > 0$.
\end{proof}

The results of Lemma \ref{lemma:s-posc-sp-properties} can be interpreted as follows. Larger values of $y_1$ and $z_1$ imply that $X_1$ is likely to take larger values. Hence, the maximum payment that buyer $1$ is willing to make in the auction stage, assuming he also knows $z_1$, is increasing in $y_1$ and nondecreasing in $z_1$. A larger value of $\alpha$ corresponds to the seller taking a larger fraction of any preliminary positive profit from the winning buyer. Buyer~$1$ compensates for this in the auction stage by lowering the maximum amount he is willing to pay; i.e., $s(\alpha, y_1, z_1)$ is decreasing in $\alpha$. Finally, inequality \eqref{eq:s-posc-sp-ineq} states that the maximum amount that buyer $1$ is willing to pay for the resource given $y_1$ and $z_1$ is no more than the expected value of the resource given $y_1$ and $z_1$, with equality holding only in the case of risk neutrality and no profit sharing in the second stage.

The strategy of a buyer is a mapping from his signal to his bid. Let buyer $n$ use strategy~$\beta_n : \mc{I}_{Y} \mapsto \R$ for $1 \leq n \leq N$. The next lemma characterizes an equilibrium bidding strategy for the second price auction with a POSC. The construction here is similar to the derivation in \cite{Milgrom&Weber82}.

\begin{lemma} \label{lemma:eq-s-posc-sp}
Let the strategies $\beta_1, \beta_2, \ldots, \beta_N$ be identical and defined by $\beta_n(y_n) \triangleq s(\alpha, y_n, y_n)$ for all~$n$. Then the strategy vector $(\beta_1, \beta_2, \ldots, \beta_N)$ is a symmetric Bayes-Nash equilibrium (BNE) of the second price auction with the POSC determined by $\alpha$. This is the unique symmetric equilibrium with an increasing and differentiable strategy.
\end{lemma}
\begin{proof}
Assume that each buyer $n > 1$ uses the strategy $\beta_n(y_n) = s(\alpha, y_n, y_n)$. We will show that the best response for buyer $1$ is to also use the strategy $\beta_1(y_1) = s(\alpha, y_1, y_1)$.

Given $y_1$, let buyer $1$ bid $b$. Buyer $1$ wins if $b \geq \max \{s(\alpha, y_n, y_n): 2 \leq n \leq N \}$.\footnote{Since the probability of a tie is zero, we ignore the issue of tie breaking without any effect on the analysis.} From Lemma~\ref{lemma:s-posc-sp-properties}, $\max\{s(\alpha, y_n, y_n): 2 \leq n \leq N\} = s(\alpha, z_1, z_1)$, where $z_1 = \max\{y_2, y_3, \ldots, y_N\}$. Thus, the expected utility of buyer $1$ is:
\begin{multline*}
\E{u\left(X_1 - s(\alpha,Z_1,Z_1) - \alpha\pos{X_1-s(\alpha,Z_1,Z_1)}\right) \indicator{b \geq s(\alpha,Z_1,Z_1)} | Y_1 = y_1} \\
= \E{\E{u\left(X_1 - s(\alpha,Z_1,Z_1) - \alpha\pos{X_1-s(\alpha,Z_1,Z_1)}\right) | Y_1 = y_1, Z_1} \indicator{b \geq s(\alpha,Z_1,Z_1)} | Y_1 = y_1}.
\end{multline*}
From \eqref{eq:s-posc-sp} and Lemma \ref{lemma:s-posc-sp-properties}, $\E{u\left(X_1 - s(\alpha,Z_1,Z_1) - \alpha\pos{X_1-s(\alpha,Z_1,Z_1)}\right) | Y_1 = y_1, Z_1}$ is positive for $Z_1 < y_1$ and negative for $Z_1 > y_1$. Thus, the expected utility is uniquely maximized by setting $b = s(\alpha, y_1, y_1)$.

Finally, consider an arbitrary symmetric equilibrium strategy vector $(\wh{\beta}, \wh{\beta}, \ldots, \wh{\beta})$ such that $ \wh{\beta}(b)$ is increasing and differentiable in $b$. We show that $s(\alpha,y_{1},y_{1})$ uniquely solves buyer $1$'s first order condition for maximizing his expected utility given the use of $\widehat{\beta}$ by all other buyers. Consequently, $\widehat{\beta}\left(  y_{n}\right) =s(\alpha,y_{n},y_{n})$ is the unique symmetric equilibrium. Define a function $\psi(y_1,z_1)$ as:
\beq{lemma-eq1-s-posc-sp}
\psi(y_1,z_1) \triangleq \CE{u\big(X_1 - \wh{\beta}(Z_1) - \alpha \posbig{X_1 - \wh{\beta}(Z_1)} \big)}{Y_1 = y_1, Z_1 = z_1 }.
\eeq
Denote the conditional pdf of $Z_1$ given $Y_1 = y_1$ by $f_{Z_1|Y_1=y_1}(z_1)$. Given $y_1$, let buyer $1$ bid $b$. The expected utility of buyer $1$ is:
\beqn
\E{\psi(Y_1,Z_1)\indicator{b \geq \wh{\beta}(Z_1)} | Y_1 = y_1} = \int_{-\infty}^{\wh{\beta}^{-1}(b)}\psi(y_1,z_1)f_{Z_1|Y_1=y_1}(t)dt.
\eeqn
The first order condition for an optimal bid $b$ gives
\begin{align} \label{eq:lemma-eq2-s-posc-sp}
&\psi(y_1,\wh{\beta}^{-1}(b))f_{Z_1|Y_1=y_1}(\wh{\beta}^{-1}(b))\od{\wh{\beta}^{-1}(b)}{b} = 0, \nonumber \\
\Rightarrow & \psi(y_1,\wh{\beta}^{-1}(b)) = 0, \nonumber \\
\Rightarrow  & \E{u\left(X_1 - b - \alpha \pos{X_1 - b} | Y_1 = y_1, Z_1 = \wh{\beta}^{-1}(b) \right)} = 0, \nonumber \\
\Rightarrow & b = s(\alpha,y_1,\wh{\beta}^{-1}(b)).
\end{align}
The second line holds because $f_{Z_1|Y_1=y_1}(t)$ is positive for all $t$ in the support of $Z_1$, and $\wh{\beta}^{-1}(b)$ is increasing in $b$ because $\wh{\beta}(b)$ is increasing in $b$. The third line is from \eqref{eq:lemma-eq1-s-posc-sp} and the last line is from~\eqref{eq:s-posc-sp}. The proof is completed by noticing that $b = \wh{\beta}(y_1)$ is an optimal bid because the strategy vector $(\wh{\beta}, \wh{\beta}, \ldots, \wh{\beta})$ constitutes a symmetric BNE.
\end{proof}

It is clear from \eqref{eq:s-posc-sp} that $s(\alpha, y_1, y_1)$ may be negative for larger values of $\alpha$. This corresponds to the case where buyer $1$ with signal $y_1$ finds the expected net gain from developing the resource after accounting for the part of the positive profit to be paid to the seller, too small to compensate for the potential loss associated with developing the resource. In such a case, buyer $1$ will not participate in the auction unless the seller pays him upon winning the auction.\footnote{We thus require that the auction be interim individually rational.} The seller, however, expects to gain from the profit sharing stage. He may thus allow the buyers to submit negative bids and make negative payments upon winning, as long as the seller's expected revenue is positive. In a second price auction with negative bids allowed, if the second highest bid is negative, by charging the winning buyer an amount equal to the second highest bid, the seller effectively pays the winning buyer in the auction stage. We show in Proposition~\ref{proposition:rev-posc-sp} that with our model assumptions, the seller's expected total revenue is always positive under this scheme. Hence, we eliminate the issue of buyers dropping out by allowing the buyers to submit negative bids.

By symmetry, the seller's expected revenues from the auction stage and from the POSC stage are the same as the expected payments made by buyer $1$ in the auction stage and in the POSC stage, respectively, conditioned on him winning the auction. In the symmetric equilibrium given by Lemma \ref{lemma:eq-s-posc-sp}, buyer $1$ wins if his signal is highest among all the buyers. Thus, the seller's expected revenue from the auction stage is $\E{s(\alpha, Z_1, Z_1) | Y_1 > Z_1}$, his expected revenue from the POSC stage is $\E{\alpha\pos{X_1-s(\alpha, Z_1, Z_1)}| Y_1 > Z_1}$, and his expected total revenue from both stages, denoted by $R^{posc}_{sp}(\alpha)$, is:
\beq{rev-eqn-posc-sp}
R^{posc}_{sp}(\alpha) \triangleq \E{s(\alpha, Z_1, Z_1) + \alpha\pos{X_1-s(\alpha, Z_1, Z_1)} | Y_1 > Z_1}.
\eeq
Taking $\alpha = 0$ corresponds to no POSC stage, i.e., the second price auction with only a one-time payment.

Proposition~\ref{proposition:rev-posc-sp} below summarizes the revenue consequences of the second price auction with a POSC. The proof is in Appendix \ref{sec:proof-rev-posc-sp}.

\begin{proposition} \label{proposition:rev-posc-sp}
The following statements hold for the second price auction with a POSC and weakly risk averse buyers:
\begin{enumerate}[(i)]
\item
The seller's expected revenue from the auction stage (possibly negative) is decreasing in the share fraction $\alpha$, while the expected revenue from the POSC stage is positive and increasing in the share fraction $\alpha$.
\item

The seller's expected total revenue from the two stages is positive and is nondecreasing in the share fraction $\alpha$; i.e., for any $0 \leq \alpha < \wh{\alpha} < 1$,
\beq{rev-posc-sp-comp}
R^{posc}_{sp}(\wh{\alpha}) \geq R^{posc}_{sp}(\alpha) > 0.
\eeq
In particular, the expected total revenue from the two stages is higher than the expected revenue from the second price auction with only a one-time payment (i.e., $R^{posc}_{sp}(0)$).
\end{enumerate}
\end{proposition}

Inequality \eqref{eq:rev-posc-sp-comp} holds even for values of $\alpha$ close to one. At these values, the second highest bid can be negative and the seller might have to pay the winning buyer in the auction stage. However, the seller extracts a large expected revenue from the POSC stage and his expected total revenue is still positive.

\subsection{Profit and loss sharing contract (PLSC)} \label{sec:plsc-sp}
Next, we consider the case where the seller takes a fraction of both any preliminary positive profit as well as any preliminary loss from the winning buyer. As in Section \ref{sec:posc-sp}, we define a function $t: [0,1) \times \mc{I}_{Y}^2 \rightarrow \R$ that will be used to characterize the bidding strategies of the buyers:
\beq{t-plsc-sp}
t(\alpha, y_1, z_1) \triangleq \left\{b: \E{u\left((1-\alpha)(X_1-b)\right) | Y_1 = y_1, Z_1 = z_1} = 0 \right\}.
\eeq
Since $u((1-\alpha)(X_1-b))$ is decreasing and continuous in~$b$, there is a unique $b$ that makes the expectation in \eqref{eq:t-plsc-sp} equal to zero. The function $t(\alpha, y_1, z_1)$ is therefore well-defined. It can be interpreted as follows. If buyer $1$ makes a payment~$b$ in the auction stage, his payment in the PLSC stage will be $\alpha(X_1 - b)$, and his total profit will be $X_1 - b - \alpha(X_1 - b) = (1-\alpha)(X_1-b)$. Thus, $t(\alpha, y_1, z_1)$ is the payment in the auction stage at which the overall expected utility of buyer $1$ is zero, conditioned on his signal being $y_1$, and the highest signal of other buyers being $z_1$.

The next lemma characterizes some important properties of $t(\alpha, y_1, z_1)$.
\begin{lemma} \label{lemma:t-plsc-sp-properties}
The function $t(\alpha, y_1, z_1)$ is increasing in $y_1$, nondecreasing in $z_1$, nondecreasing in $\alpha$ (increasing if $u$ is strictly concave), and positive everywhere. Moreover, for all $y_1$, $z_1$, and $\alpha \in [0,1)$,
\beq{t-plsc-sp-ineq}
s(\alpha, y_1, z_1) \leq t(\alpha, y_1, z_1) \leq \E{X_1 | Y_1 = y_1, Z_1 = z_1},
\eeq
where $s(\alpha, y_1, z_1)$ is defined by \eqref{eq:s-posc-sp}. The left inequality above is strict everywhere except for $\alpha = 0$, and the right inequality is strict if $u$ is strictly concave.
\end{lemma}
\begin{proof}
From Lemma \ref{lemma:fosd-sp}, $\E{u((1-\alpha)(X_1-b)) | Y_1 = y_1, Z_1 = z_1}$ is increasing in $y_1$ and nondecreasing in $z_1$. Then \eqref{eq:t-plsc-sp} implies that $t(\alpha, y_1, z_1)$ is increasing in $y_1$ and nondecreasing in $z_1$. Next, let $\alpha$ and~$\wh{\alpha}$ be such that $0 \leq \alpha < \wh{\alpha} < 1$. Since $u$ is concave,
\begin{align} \label{eq:t-plsc-sp-properties-eq1}
u((1-\wh{\alpha})(X_1 - t(\alpha,y_1,z_1))) 
& = u\left(\left(\frac{1-\wh{\alpha}}{1-\alpha}\right)(1-\alpha)(X_1 - t(\alpha,y_1,z_1)) \right), \nonumber \\ 
& \geq  \left(\frac{1-\wh{\alpha}}{1-\alpha}\right)u\left((1-\alpha)(X_1 - t(\alpha,y_1,z_1))\right) + \left(1- \frac{1-\wh{\alpha}}{1-\alpha}\right)u(0), \nonumber \\ 
& = \left(\frac{1-\wh{\alpha}}{1-\alpha}\right)u\left((1-\alpha)(X_1 - t(\alpha,y_1,z_1))\right),
\end{align}
where the inequality is strict if $u$ is strictly concave. From \eqref{eq:t-plsc-sp} and \eqref{eq:t-plsc-sp-properties-eq1},
\begin{multline*}
\E{u((1-\wh{\alpha})(X_1 - t(\alpha,y_1,z_1))) | Y_1 = y_1, Z_1 = z_1} \\
\geq \left(\frac{1-\wh{\alpha}}{1-\alpha}\right)\E{u((1-\alpha)(X_1 - t(\alpha,y_1,z_1))) | Y_1 = y_1, Z_1 = z_1} = 0.
\end{multline*}
$\E{u((1-\alpha)(X_1-b)) | Y_1 = y_1, Z_1 = z_1}$ is a decreasing function of $b$. Then from \eqref{eq:t-plsc-sp}, we must have $t(\wh{\alpha},y_1,z_1) \geq t(\alpha,y_1,z_1)$. This shows that $t(\alpha, y_1, z_1)$ is nondecreasing in $\alpha$ (increasing in $\alpha$ if $u$ is strictly concave).

From \eqref{eq:s-posc-sp} and \eqref{eq:t-plsc-sp}, $s(0,y_1,z_1) = t(0,y_1,z_1)$. Since $s$ is decreasing in $\alpha$ while $t$ is nondecreasing in $\alpha$, we get $t(\alpha,y_1,z_1) \geq t(0,y_1,z_1) =  s(0,y_1,z_1) > 0$. Hence, $t(\alpha,y_1,z_1)$ is positive everywhere. Moreover, $t(\alpha,y_1,z_1) \geq s(\alpha,y_1,z_1)$, where the equality holds only at $\alpha = 0$. Finally, Jensen's inequality and \eqref{eq:t-plsc-sp} imply
\begin{multline*}
u((1-\alpha)(\E{X_1 | Y_1 = y_1, Z_1 = z_1} - t(\alpha,y_1,z_1)))  \\ \geq \E{u((1-\alpha)(X_1 - t(\alpha,y_1,z_1))) | Y_1 = y_1, Z_1 = z_1} = 0,
\end{multline*}
which establishes the second inequality (strict inequality if $u$ is strictly concave).
\end{proof}

Recall from Lemma \ref{lemma:s-posc-sp-properties} that the payment $s(\alpha,y_1,z_1)$ that makes buyer $1$ indifferent to winning in a POSC given $y_1$ and $z_1$ is nonincreasing in the share fraction $\alpha$; buyer $1$ can only be willing to pay less as the seller takes a larger share in the POSC stage. The corresponding payment $t(\alpha,y_1,z_1)$ in the PLSC, however, can only remain constant or increase as the share fraction $\alpha$ increases. This is because a larger $\alpha$ not only reduces his preliminary positive profits but also shields him from preliminary losses. If buyers are strictly risk averse, then a larger $\alpha$ allows buyer $1$ to make a strictly larger payment, reflecting the additional benefit that he receives from transferring risk to the seller.

As in Section \ref{sec:posc-sp}, let $\beta_n$ denote the strategy of buyer $n$, for each $n$. The next lemma characterizes an equilibrium bidding strategy for the second price auction with a PLSC. 

\begin{lemma} \label{lemma:eq-t-plsc-sp}
Let the strategy of a buyer $n$ be $\beta_n(y_n) = t(\alpha, y_n, y_n)$. Then the strategy vector $(\beta_1, \beta_2, \ldots, \beta_N)$ is a symmetric BNE of the second price auction with the PLSC determined by $\alpha$. This equilibrium is the unique symmetric equilibrium with an increasing and differentiable strategy.
\end{lemma}
\begin{proof}
The proof is similar to that of Lemma \ref{lemma:eq-s-posc-sp} and is obtained by using the function $t(\alpha, y_1, z_1)$ instead of the function $s(\alpha, y_1, z_1)$.
\end{proof}

The equilibrium bid $t(\alpha, y_n, y_n)$ in a PLSC is always positive. Unlike a POSC, the issue of buyers either dropping out or the seller allowing the buyers to submit negative bids does not arise here.

Again, by symmetry, the seller's expected revenues from the auction stage and from the PLSC stage are the same as the expected payments made by buyer $1$ in the auction stage and in the PLSC stage, respectively, conditioned on him winning the auction. In the symmetric equilibrium given by Lemma \ref{lemma:eq-t-plsc-sp}, buyer $1$ wins if his signal is highest among all the buyers. Thus, the seller's expected revenue from the auction stage is $\E{t(\alpha, Z_1, Z_1) | Y_1 > Z_1}$, his expected revenue from the PLSC stage is $\E{\alpha(X_1-t(\alpha, Z_1, Z_1))| Y_1 > Z_1}$, and his expected total revenue from both stages, denoted by $R^{plsc}_{sp}(\alpha)$, is:
\beq{rev-eqn-plsc-sp}
R^{plsc}_{sp}(\alpha) \triangleq \E{\alpha X_1 + (1-\alpha) t(\alpha, Z_1, Z_1) | Y_1 > Z_1}.
\eeq
Again, $\alpha = 0$ corresponds to the second price auction with only a one-time payment, and so $R^{plsc}_{sp}(0) = R^{posc}_{sp}(0)$.

Proposition~\ref{proposition:rev-plsc-sp} below summarizes the revenue consequences of the second price auction with a PLSC. The proof is in Appendix \ref{sec:proof-rev-plsc-sp}.

\begin{proposition} \label{proposition:rev-plsc-sp}
The following statements hold for the second price auction with a PLSC and weakly risk averse buyers:
\begin{enumerate}[(i)]
\item
The seller's expected revenue from the auction stage is positive and is nondecreasing in the share fraction $\alpha$. The expected revenue from the PLSC stage is also positive. 
\item
The seller's expected total revenue from the two stages is positive and is increasing in the share fraction $\alpha$; i.e., for any $0 \leq \alpha < \wh{\alpha} < 1$,
\beq{rev_plsc-sp-comp}
R^{plsc}_{sp}(\wh{\alpha}) > R^{plsc}_{sp}(\alpha).
\eeq
In particular, the expected total revenue from the two stages is higher than the expected revenue from the second price auction with only a one-time payment (i.e., $R^{plsc}_{sp}(0)$).
\end{enumerate}
\end{proposition}

A natural question now is: How does the second price auction with a PLSC compare with the second price auction with a POSC in terms of the expected total revenue they generate for the same share fraction $\alpha$? The two procedures are identical at $\alpha = 0$. For $\alpha \in (0,1)$, since $t(\alpha, Z_1, Z_1) > s(\alpha, Z_1, Z_1)$, the seller's expected revenue from the auction stage is higher in the case of a PLSC than a POSC. However, $\alpha (X_1 - t(\alpha, Z_1, Z_1)) < \alpha \pos{X_1 - s(\alpha, Z_1, Z_1)}$. Thus, the seller's expected revenue from the profit sharing stage is higher in the case of a POSC than a PLSC. As a result, it is not immediately clear how the expected total revenue from the two stages in the PLSC compares with the expected total revenue from the POSC. The next proposition resolves this issue. It shows that in the PLSC, the gain from the higher bidding in the auction stage outweighs the loss in revenue in the PLSC stage. The proof is in Appendix \ref{sec:proof-rev-plsc-posc-sp}. 

\begin{proposition} \label{proposition:rev-plsc-posc-sp}
The second price auction with a PLSC and weakly risk averse buyers generates higher expected total revenue than the second price auction with a POSC and weakly risk averse buyers; i.e., for any $\alpha \in (0,1)$,
\beq{rev-plsc-posc-sp-comp}
R_{sp}^{plsc}(\alpha) \geq R_{sp}^{posc}(\alpha).
\eeq
\end{proposition}

\subsection{Revenue ranking for risk neutral buyers} \label{sec:rn-sp}

The seller is risk neutral in our model while the buyers are weakly risk averse. The return to the winning buyer is less risky in the PLSC determined by $\alpha>0$ than in the corresponding POSC. In the case of strictly risk averse buyers, the PLSC thus creates more potential gains from risk sharing than the POSC. This suggests that the ranking \eqref{eq:rev-plsc-posc-sp-comp} reflects a successful effort by the seller to garner a portion of these gains in the form of higher revenue by using a PLSC. We explain below, however, why \eqref{eq:rev-plsc-posc-sp-comp} holds even in the case of risk neutral buyers as a consequence of the form of positive dependence among signals and return assumed in Section \ref{sec:model}, and the nature of PLSC and POSC. This is followed by an example in which \eqref{eq:rev-plsc-posc-sp-comp} holds strictly in the case of risk neutral buyers. The effect of risk sharing on revenue is thus aligned with but not the primary cause of this ranking.

Fix a share fraction $\alpha > 0$ and consider $y_1 > z_1$, i.e., buyer $1$ wins in both the POSC and the PLSC. Let $r^{posc}(x_1)$ and $r^{plsc}(x_1)$ be the ex-post total revenue to the seller from the POSC and the PLSC respectively, if the value $X_1$ of the resource is equal to $x_1$:
\begin{align}
r^{posc}(x_1) & \triangleq s(\alpha, z_1, z_1) + \alpha \pos{x_1 - s(\alpha, z_1, z_1)}, \label{eq:rev-rn-posc} \\
r^{plsc}(x_1) & \triangleq t(\alpha, z_1, z_1) + \alpha (x_1 - t(\alpha, z_1, z_1)). \label{eq:rev-rn-plsc} 
\end{align}
Buyer $1$ in each case pays in the auction stage the bid of the buyer who observed $z_1$. This is equal to $s(\alpha,z_{1},z_{1})$ in the POSC and $t(\alpha,z_{1},z_{1})$ in the PLSC. In our symmetric model with risk neutral buyer (i.e., $u(x) = x$), $s(\alpha, z_1, z_1)$ and $t(\alpha, z_1, z_1)$ are the bids that would make buyer $1$ indifferent to winning conditioned on $Y_1 = Z_1 = z_1$ in the POSC and the PLSC, respectively:
\begin{align}
& \CE{X_1 - s(\alpha, z_1, z_1) -\alpha \pos{X_1 - s(\alpha, z_1, z_1)}}{Y_1 = z_1, Z_1 =z_1} = 0, \label{eq:rn-s} \\
& \CE{X_1 - t(\alpha, z_1, z_1) -\alpha (X_1 - t(\alpha, z_1, z_1))}{Y_1 = z_1, Z_1 =z_1} = 0. \label{eq:rn-t}
\end{align}

Equations \eqref{eq:rn-s} and \eqref{eq:rn-t} together imply that the difference $r^{plsc}(x_{1})-r^{posc}(x_{1})$ satisfies:
\beq{rn-st} 
\CE{r^{plsc}(X_1) - r^{posc}(X_1)}{Y_1 = z_1, Z_1 = z_1} = 0.
\eeq
The seller would thus expect to receive the same revenue from the PLSC as in the POSC if the highest and the second highest signals are the same, i.e.,  $Y_1=Z_1=z_1$. The event in which he sells to buyer $1$, however, is defined by $Y_1>Z_1$. The difference $r^{plsc}(x_1)-r^{posc}(x_1)$ is a line with slope $\alpha$ for $x_1 <s(\alpha,z_1,z_1)$ and constant for $x_1 \geq s(\alpha,z_1,z_1)$; it is therefore nondecreasing for all $x_1$ and increasing for $x_1<s(\alpha,z_1,z_1)$. It can then be shown using Lemma~\ref{lemma:fosd-sp} the left side of~\eqref{eq:rn-st} is nondecreasing in buyer $1$'s signal $Y_{1}$; i.e.,
\beqn 
\CE{r^{plsc}(X_1) - r^{posc}(X_1)}{Y_1 = y_1, Z_1 = z_1} \geq 0
\eeqn
for $y_{1}>z_{1}$. The PLSC with share fraction $\alpha$ therefore generates a higher expected revenue than the corresponding POSC.

\begin{example} \label{eg:rev-ranking}
Consider two risk neutral buyers. The signals $Y_1$ and $Y_2$ are independent and uniformly distributed in $(0,1)$. Given the signals of the buyers, the value of the resource to buyer $1$ and buyer $2$ are Bernoulli random variables~$X_1$ and $X_2$ respectively, with $\Prob{X_1 = 1| Y_1 = y_1, Y_2 = y_2} = (2y_1 + y_2)/3$ and $\Prob{X_2 = 1| Y_1 = y_1, Y_2 = y_2} = (y_1 + 2y_2)/3$. Clearly, Assumptions \ref{assumption:symmetry} and \ref{assumption:pos-dep} are satisfied. Also, $Z_1 = Y_2$.
 
For the second price auction with the POSC determined by $\alpha$, the function $s(\alpha,y_1,z_1)$ is obtained as follows:
\begin{align*}
s(\alpha, y_1, z_1) 
& = \left\{b : \E{X_1-b-\alpha\pos{X_1-b}|Y_1 = y_1, Z_1 = z_1} = 0 \right\}, \\
& = \left\{b : \frac{2y_1+z_1}{3} - b - \alpha(1-b)\frac{2y_1+z_1}{3} = 0 \right\}, \\
& = \frac{(1-\alpha)(2y_1 + z_1)}{3-\alpha(2y_1+z_1)}.
\end{align*}
Thus, buyer $1$ bids $(1-\alpha)y_1/(1-\alpha y_1)$ if his signal is $y_1$ and buyer $2$ bids $(1-\alpha)y_2/(1-\alpha y_2)$ if his signal is $y_2$. The seller's expected revenue from the auction stage is:
\beqn
\E{\frac{(1-\alpha)\min(Y_1, Y_2)}{1-\alpha \min(Y_1, Y_2)}} = 2(1-\alpha)\int_0^1 \frac{\theta(1-\theta)}{1-\alpha\theta}d\theta.
\eeqn
Here, we have used the fact that the pdf of $\min(Y_1, Y_2)$ is $2(1-\theta)$ for $\theta \in (0,1)$. The seller's expected revenue from the POSC stage is:
\beqn
\E{\alpha\pos{X_1 - \frac{(1-\alpha)Y_2}{1-\alpha Y_2}} \bigg| Y_1 > Y_2} 
= 2\alpha \E{\frac{2Y_1 + Y_2}{3} \left(1 - \frac{(1-\alpha)Y_2}{1-\alpha Y_2}\right)\indicator{Y_1 > Y_2}}.
\eeqn
Since $\E{(2Y_1 + Y_2)\indicator{Y_1 > Y_2} | Y_2} = (1-Y_2)(1+2Y_2)$, the above expression can be further simplified to
\beqn
\frac{5\alpha}{9} - \frac{2\alpha(1-\alpha)}{3}
\E{\frac{Y_2(1-Y_2)(1+2Y_2)}{1-\alpha Y_2}} 
= \frac{5\alpha}{9} - \frac{2\alpha(1-\alpha)}{3}
\int_0^1\frac{\theta(1-\theta)(1+2\theta)}{1-\alpha\theta}d\theta.
\eeqn
With some further (tedious) algebra, one can get closed form expressions for the seller's expected revenue from the auction stage and the seller's revenue from the POSC stage. 

For the second price auction with the PLSC determined by $\alpha$, the function $t(\alpha,y_1,z_1)$ is obtained as follows:
\begin{align*}
t(\alpha, y_1, z_1) 
& = \left\{b : \E{(1-\alpha)(X_1-b)|Y_1 = y_1, Z_1 = z_1} = 0 \right\}, \\
& = \frac{2y_1 + z_1}{3}.
\end{align*}
As a consequence, each buyer bids his signal. The seller's expected revenue from the auction stage is:
\beqn
\E{\min(Y_1, Y_2)} = \frac{1}{3}.
\eeqn
The seller's expected revenue from the PLSC stage is:
\beqn
\E{\alpha(X_1 - Y_2) | Y_1 > Y_2} = 2\alpha\E{\left(\frac{2Y_1 + Y_2}{3}-Y_2\right)\indicator{Y_1 > Y_2}} = \frac{2\alpha}{9}.
\eeqn

\begin{figure}[t]
\begin{center}
\includegraphics[trim=0.5in 0.35in 0.5in 0.3in, clip=true, height=4.5in]{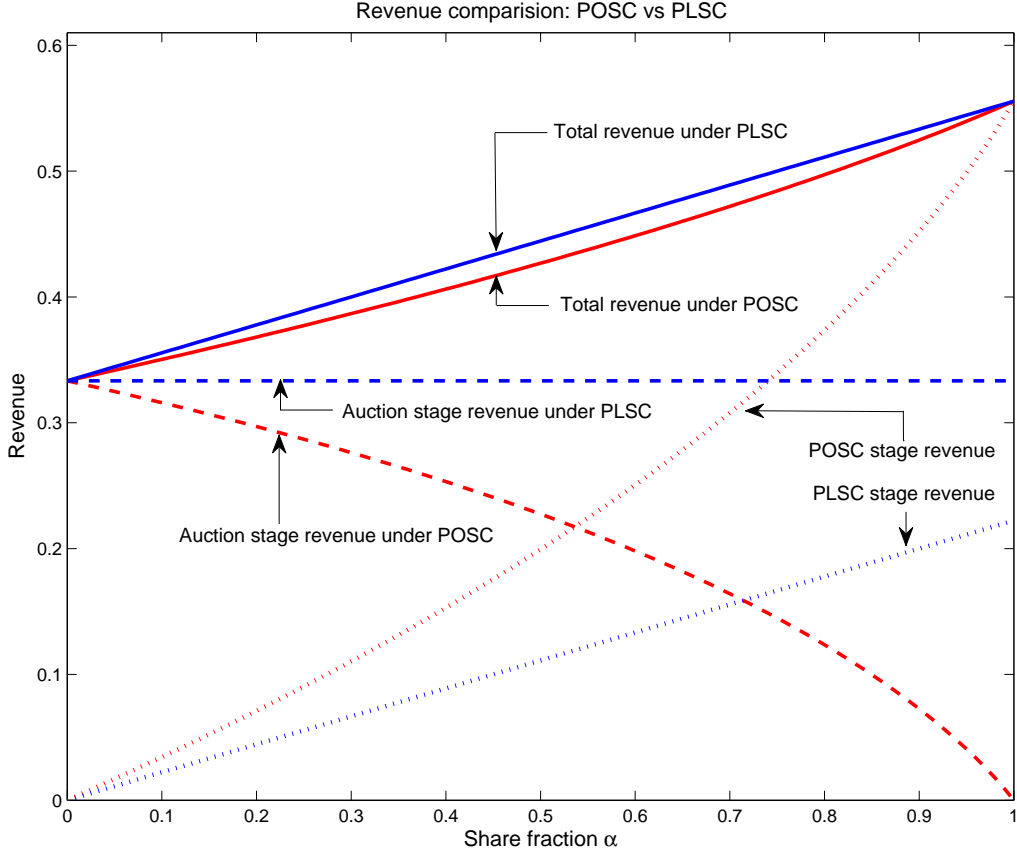}
\caption{\small \sl Revenue from the second price auction with a POSC and a PLSC as a function of the share fraction $\alpha$. \label{fig:posc-vs-plsc-plot}} 
\end{center} 
\end{figure}

The results are plotted in Figure~\ref{fig:posc-vs-plsc-plot}. For POSC, it can be seen that as $\alpha$ increases, the expected revenue from the auction stage decreases while the revenue from the POSC stage increases. However, the expected total revenue is increasing in $\alpha$. In particular, the expected total revenue with $\alpha > 0$ is higher than the expected total revenue with $\alpha = 0$, i.e., the second price auction with only a one-time payment. For PLSC, the expected revenue from the auction stage and the expected total revenue are increasing in $\alpha$. The expected total revenue from the second price auction with a PLSC is strictly higher than the expected total revenue from the second price auction with a POSC; they agree only at $\alpha = 0$, the case of only a one-time payment, and in the limiting case of $\alpha \rightarrow 1$, the case of full profit extraction.

\end{example}

\section{The English Auction with a Profit Sharing Contract} \label{sec:english}
This section extends the results of Section \ref{sec:second-price} to the English auction. In an English auction, the prices at which different buyers drop out provide information to the remaining buyers. A buyer's bidding strategy takes into account the number of active buyers and the prices at which the other buyers have dropped out. Again, we look for a symmetric equilibrium.

The following lemma is a consequence of Assumption \ref{assumption:pos-dep}. It plays a role similar to that of Lemma~\ref{lemma:fosd-sp} in Section \ref{sec:second-price}.
\begin{lemma} \label{lemma:fosd-eng}
$\E{h(X_1)|Y_1 = y_1, \mb{Z} = \mb{z}}$ is increasing in $y_1$ and nondecreasing in $\mb{z}$ (componentwise) for any increasing function $h:\R \mapsto \R$ for which the expectation exists.
\end{lemma}
\begin{proof}
Recall the definition of \mb{Z} from section \ref{sec:model}. The proof follows trivially from Assumptions \ref{assumption:symmetry} and \ref{assumption:pos-dep}.
\end{proof}

We now analyze the English auction with a POSC and a PLSC. Paralleling Section \ref{sec:posc-sp} and Section \ref{sec:plsc-sp}, we (re)define functions $s: [0,1) \times \mc{I}_{Y}^N \rightarrow \R$ and $t: [0,1) \times \mc{I}_{Y}^N \rightarrow \R$ that will be used to characterize the bidding strategy of buyer $1$:
\begin{align}
s(\alpha, y_1, \mb{z}) & \triangleq \left\{b: \E{u\left( X_1 - b - \alpha\pos{X_1 - b} \right) | Y_1 = y_1, \mb{Z} = \mb{z}} = 0 \right\}. \label{eq:s-posc-eng} \\
t(\alpha, y_1, \mb{z}) & \triangleq \left\{b: \E{u\left((1-\alpha)(X_1-b)\right) | Y_1 = y_1, \mb{Z} = \mb{z}} = 0 \right\}. \label{eq:t-plsc-eng}
\end{align}
The functions $s(\alpha, y_1, \mb{z})$ and $t(\alpha, y_1, \mb{z})$ are well-defined for reasons similar to those in Sections~\ref{sec:posc-sp} and~\ref{sec:plsc-sp} respectively. These are the maximum payments that buyer $1$ is willing to pay in the auction stage under the POSC and PLSC, respectively, given that his signal is $y_1$ and the signals of everyone else is~$\mb{z}$ (the assignment of these signals to the buyers is irrelevant). The next lemma characterizes some important properties of $s(\alpha, y_1, \mb{z})$ and $t(\alpha, y_1, \mb{z})$. The results are fairly intuitive and explanations similar to those for Lemma \ref{lemma:s-posc-sp-properties} and Lemma~\ref{lemma:t-plsc-sp-properties} apply.

\begin{lemma} \label{lemma:st-eng-properties}
The function $s(\alpha, y_1, \mb{z})$, defined by \eqref{eq:s-posc-eng}, is increasing in $y_1$, nondecreasing in $\mb{z}$, decreasing in~$\alpha$, and positive for small values of $\alpha$. The function $t(\alpha, y_1, \mb{z})$, defined by \eqref{eq:t-plsc-eng}, is increasing in $y_1$, nondecreasing in $\mb{z}$, nondecreasing in $\alpha$ (increasing if $u$ is strictly concave), and is always positive. Moreover, for all $y_1$, $\mb{z}$, and $\alpha \in [0,1)$,
\beq{st-eng-ineq}
s(\alpha, y_1, \mb{z}) \leq t(\alpha, y_1, \mb{z}) \leq \E{X_1 | Y_1 = y_1, \mb{Z} = \mb{z}},
\eeq
where the left inequality is strict everywhere except for $\alpha = 0$, and the right inequality is strict if $u$ is strictly concave.
\end{lemma}
\begin{proof}
The properties of $s(\alpha, y_1, \mb{z})$ follow from Lemma \ref{lemma:fosd-eng}, \eqref{eq:s-posc-eng}, and concavity of $u$. The sequence of arguments is similar to how Lemma \ref{lemma:fosd-sp}, \eqref{eq:s-posc-sp}, and concavity of $u$ are used to prove Lemma~\ref{lemma:s-posc-sp-properties}. The properties of $t(\alpha, y_1, \mb{z})$ follow from Lemma \ref{lemma:fosd-eng}, \eqref{eq:t-plsc-eng}, and concavity of $u$. The sequence of arguments is similar to how Lemma \ref{lemma:fosd-sp}, \eqref{eq:t-plsc-sp}, and concavity of $u$ are used to prove Lemma~\ref{lemma:t-plsc-sp-properties}.
\end{proof}

In an English auction, the price of the resource increases continuously during the bidding stage. At each price level, a buyer decides whether to drop out or not, depending on his signal and the prices at which other buyers have dropped out. Let $p_1 \leq p_2 \leq \ldots \leq p_{N-1}$ be the prices at which the first, second, $\ldots$, last drop out occurs. The strategy of a buyer~$n$ now is a vector $\bs{\beta}_n \triangleq (\beta_n^2, \beta_n^3, \ldots, \beta_n^N)$ of functions $\beta_n^k : \R^{N-k+1} \rightarrow \R$, $2 \leq k \leq N$. Here, $\beta_n^k(y_n, p_{N-k}, p_{N-k-1}, \ldots, p_{1})$ is the price at which buyer~$n$ drops out as a function of his signal $y_n$ and the prices $(p_{N-k}, p_{N-k-1}, \ldots, p_{1})$ at which the $N-k$ drop outs have occurred. The next lemma characterizes an equilibrium bidding strategy for the English auction with a POSC and a PLSC. The construction is similar to how an equilibrium bidding strategy for the English auction is characterized in \cite{Milgrom&Weber82}. 

\begin{lemma} \label{lemma:eq-st-eng}
Let $p_1 \leq p_2 \leq \ldots \leq p_{N-1}$ be the prices at which buyers drop out. Define $q_1 \leq q_2 \leq \ldots \leq q_{N-1}$ recursively as follows:
\begin{align} \label{q-recursion-posc}
q_1 & \triangleq \Big\{b: s\big(\alpha, \underbrace{b, b, \ldots, b}_{\text{N times}}\big) = p_1 \Big\}, \nonumber \\
q_k & \triangleq \Big\{b: s\big(\alpha, \underbrace{b, b, \ldots, b}_{\text{N-k+1 times}}, q_{k-1}, q_{k_2}, \ldots, q_1 \big) = p_k \Big\}, \quad \text{for $2 \leq k \leq N-1$}.
\end{align}
Let the strategy of each buyer $n$ be defined recursively as follows:
\begin{align} \label{eq:bid-recursion-posc}
\beta_n^N(y_n) & = s\big(\alpha, \underbrace{y_n, y_n, \ldots, y_n}_{\text{N times}}\big), \nonumber \\
\beta_n^k(y_n, p_{N-k}, p_{N-k-1}, \ldots, p_{1}) & = s\big(\alpha, \underbrace{y_n, y_n, \ldots, y_n}_{\text{k times}},  q_{N-k}, q_{N-k-1}, \ldots, q_{1} \big) \quad \text{for $2 \leq k \leq N-1$}.
\end{align} 
Then $(\bs{\beta}_1 , \bs{\beta}_2 , \ldots, \bs{\beta}_N )$ is a symmetric perfect Bayesian equilibrium (PBE) of the English auction with the POSC determined by $\alpha$. A symmetric PBE of the English auction with the PLSC determined by $\alpha$ is characterized similarly by defining the $q_k$'s and the $\beta_n^k$'s using $t(\alpha, y_1, \mb{z})$.
\end{lemma}
\begin{proof}
We only give the proof for the English auction with a POSC. The proof for the English auction with a PLSC is similar.

Assume that each buyer $n >1$ use the strategy $\bs{\beta}_n$. We will show that an optimal response of buyer $1$ is to use the strategy $\bs{\beta}_1$. Notice that the $q_k$'s are well-defined as $s(\alpha, y_1, \mb{z})$ is increasing in~$y_1$ and nondecreasing in $\mb{z}$.

Since each buyer $n>1$ uses the same increasing strategy $\bs{\beta}_n$, if buyer $1$ wins the auction, then it is straightforward to verify that $q_k = z_{N-k}$ for $1 \leq k \leq N-1$, where $z_{N-k}$ is the $(N-k)^{th}$ highest signal from the signals of the buyers $2, 3, \ldots, N$. Thus, the $q_k$'s are inverse mappings that compute the signals of the buyers who drop out from the prices at which they do so. The price that buyer $1$ pays upon winning is equal to $s(\alpha, z_1, \mb{z})$. Buyer $1$ cannot influence the price he pays if he wins the auction. From \eqref{eq:s-posc-eng} and Lemma \ref{lemma:st-eng-properties}, the expected utility of buyer $1$ is positive if $y_1 > z_1$ (here, $s(\alpha, y_1, \mb{z}) > s(\alpha, z_1, \mb{z})$) and negative if $y_1 < z_1$ (here, $s(\alpha, y_1, \mb{z}) < s(\alpha, z_1, \mb{z})$).

Suppose $y_1 > z_1$. Any strategy by which buyer $1$ wins gives him the same positive expected utility, while the utility is zero if he drops out. In particular, if buyer $1$ uses the strategy $\bs{\beta}_1$, he wins the auction. Hence, $\bs{\beta}_1$ is optimal if $y_1 > z_1$. Next, if $y_1 < z_1$, then any strategy by which buyer $1$ wins gives him the same negative expected utility, while the utility is zero if he drops out. By following the strategy $\bs{\beta}_1$, buyer $1$ drops out; more specifically, if his signal is the $l^{th}$ highest, where $l \geq 2$, then buyer $1$ will be $(N-l)^{th}$ to drop out. Thus, $\bs{\beta}_1$ is optimal if $y_1 < z_1$. This completes the proof.
\end{proof}

As in Section \ref{sec:posc-sp}, we assume in the case of POSC that the seller allows the buyers to submit negative bids and always charges the winning buyer the price at which the second to last buyer drops out, even if it is negative. By contrast, the equilibrium bids in the PLSC are always positive.

The English auction has been shown to have infinitely many symmetric equilibria in the model of \cite{Milgrom&Weber82}; see, e.g., \cite{Bikhchandani2002}. However, all these equilibria are equivalent in the sense that they result in a two-stage procedure: the buyers with the lowest $N-2$ signals drop out and reveal their signals in this process and then the last two buyers carry out the second price auction while taking into account the lowest $N-2$ signals. In particular, all such equilibria agree on the price at which the second to last buyer drops out, hence the seller's revenue does not depend on the choice of the equilibrium. A similar result, on the lines of \cite{Bikhchandani2002}, can be obtained for the English auction with a POSC and a PLSC in our model. We omit the details here.

The seller's expected revenues from the auction stage and from the profit sharing stage are same as the expected payments made by buyer $1$ in the auction stage and in the profit sharing stage, respectively, conditioned on him winning the auction. In the symmetric equilibrium given by Lemma \ref{lemma:eq-st-eng}, buyer $1$ wins if his signal is highest among all the buyers. In a POSC with the share fraction $\alpha$, the seller's expected revenue from the auction stage is $\E{s(\alpha, Z_1, \mb{Z}) | Y_1 > Z_1}$, his expected revenue from the POSC stage is $\E{\alpha\pos{X_1-s(\alpha, Z_1, \mb{Z})}| Y_1 > Z_1}$, and his expected total revenue from both stages, denoted by $R^{posc}_{eng}(\alpha)$, is:
\beq{rev-eqn-posc-eng}
R^{posc}_{eng}(\alpha) \triangleq \E{s(\alpha, Z_1, \mb{Z}) + \alpha\pos{X_1-s(\alpha, Z_1, \mb{Z})} | Y_1 > Z_1},
\eeq 
where $s(\alpha, y_1, \mb{z})$ is given by \eqref{eq:s-posc-eng}. In a PLSC with the share fraction $\alpha$, the seller's expected revenue from the auction stage is $\E{t(\alpha, Z_1, \mb{Z}) | Y_1 > Z_1}$, his expected revenue from the PLSC stage is $\E{\alpha(X_1-t(\alpha, Z_1, \mb{Z}))| Y_1 > Z_1}$, and his expected total revenue from both stages, denoted by $R^{plsc}_{eng}(\alpha)$, is:
\beq{rev-eqn-plsc-eng}
R^{plsc}_{eng}(\alpha) \triangleq \E{\alpha X_1 + (1-\alpha)t(\alpha, Z_1, \mb{Z}) | Y_1 > Z_1},
\eeq
where $t(\alpha, y_1, \mb{z})$ is defined by \eqref{eq:t-plsc-eng}. Taking $\alpha = 0$ corresponds to the English auction with only a one-time payment, and so $R^{plsc}_{eng}(0) = R^{posc}_{eng}(0)$.

Proposition~\ref{proposition:rev-eng} below extends the results of Propositions~\ref{proposition:rev-posc-sp}-\ref{proposition:rev-plsc-posc-sp} to the English auction.

\begin{proposition} \label{proposition:rev-eng}
The following statements hold in an English auction with weakly risk averse buyers:
\begin{enumerate}[(i)]
\item
In the POSC, the seller's expected revenue from the auction stage (possibly negative) is decreasing in the share fraction $\alpha$; the expected revenue from the POSC stage is positive and increasing in the share fraction $\alpha$; and the expected total revenue from the two stages is positive and nondecreasing in the share fraction $\alpha$, i.e., for any $0 \leq \alpha < \wh{\alpha} < 1$,
\beq{rev-posc-eng-comp}
R^{posc}_{eng}(\wh{\alpha}) \geq R^{posc}_{eng}(\alpha) > 0.
\eeq

\item
In the PLSC, the seller's expected revenue from the auction stage is positive and increasing in the share fraction $\alpha$; the expected revenue from the PLSC stage is positive; and the expected total revenue from the two stages is positive and increasing in the share fraction $\alpha$, i.e., for any $0 \leq \alpha < \wh{\alpha} < 1$,
\beq{rev_plsc-eng-comp}
R^{plsc}_{eng}(\wh{\alpha}) > R^{plsc}_{eng}(\alpha).
\eeq

\item
The PLSC with a share fraction $\alpha$ generates higher expected total revenue than the POSC with a share fraction $\alpha$; i.e., for any $\alpha \in (0,1)$,
\beq{rev-plsc-posc-eng-comp}
R_{eng}^{plsc}(\alpha) \geq R_{eng}^{posc}(\alpha).
\eeq
\end{enumerate}
In particular, the expected total revenue from the two stages in the POSC or PLSC is higher than the expected revenue from the English auction with only a one-time payment.
\end{proposition}
\begin{proof}
The proof of the first part of the claim follows a sequence of arguments similar to that in the proof of Proposition~\ref{proposition:rev-posc-sp}. The necessary intermediate steps are now obtained by using Lemma~\ref{lemma:fosd-eng},~\eqref{eq:s-posc-eng}, and the concavity of~$u$; paralleling the use of Lemma~\ref{lemma:fosd-sp},~\eqref{eq:s-posc-sp}, and the concavity of~$u$ to prove Proposition~\ref{proposition:rev-posc-sp}. The proof of the second part of the claim follows a sequence of arguments similar to that in the proof of Proposition~\ref{proposition:rev-plsc-sp}. The necessary intermediate steps are now obtained by using Lemma~\ref{lemma:fosd-eng} and~\eqref{eq:t-plsc-eng}, paralleling the use of Lemma~\ref{lemma:fosd-sp} and~\eqref{eq:t-plsc-sp} to prove Proposition~\ref{proposition:rev-plsc-sp}. The proof of the third part of the claim follows a sequence of arguments similar to that in the proof of Proposition~\ref{proposition:rev-plsc-posc-sp}. The necessary intermediate steps are now obtained by using Lemma~\ref{lemma:fosd-eng},~\eqref{eq:s-posc-eng},~\eqref{eq:t-plsc-eng}, and the concavity of~$u$; paralleling the use of Lemma~\ref{lemma:fosd-sp},~\eqref{eq:s-posc-sp},~\eqref{eq:t-plsc-sp}, and the concavity of~$u$ to prove Proposition~\ref{proposition:rev-plsc-posc-sp}.
\end{proof}

\section{Discussion} \label{sec:discussion}
This section describes some extensions of the model and results of the previous sections. \newline

\noindent \textbf{Discounting and multiperiod profit realization:} 
Our results for POSC and PLSC extend directly to the case in which the seller and each buyer discounts the return from the resource relative to the auction stage payment using a common discount factor. If the return from the resource is generated over multiple periods, then our results extend for the PLSC by substituting the discounted sum of per period return for the variable $X_1$ in our analysis. A subtlety arises in the multiple period case for the POSC: Is the winning buyer required to share a positive profit earned period by period, or is he instead required to share only based upon the net discounted profit over all of the periods? The distinction is simply that a positive profit in one period may be canceled by a loss in a subsequent period. Our analysis extends in the case of POSC if a buyer is taxed based upon discounted net profit; i.e., the winning buyer can apply losses in some periods to cancel profits in other periods, as is common in tax law.\footnote{For $1 \leq t \leq T$, let the random variable $X_n(t)$ be the return from the resource to buyer $n$ in time period $t$ and $\delta \in (0,1]$ be the common discount factor. If buyer $1$ wins the auction, makes a payment $b$ in the auction stage, and reports a preliminary profit $X_1(t) - b(t)$ in time period $t$, then the $b(t)$'s must satisfy $b = \sum_{t=1}^T \delta^t b(t)$. The total profit of buyer $1$ from $T$ time periods, relative to the auction stage payment, is $(1-\alpha)\big(\sum_{t=1}^T \delta^t X_1(t) - b \big)$ in the PLSC and is $\sum_{t = 1}^T \alpha \delta^t \pos{X_1(t) - b(t)}$ in the POSC with period by period profit sharing. The subtlety with the POSC arises because $\sum_{t = 1}^T \alpha \delta^t \pos{X_1(t) - b(t)} \geq \alpha [\sum_{t = 1}^T \delta^t X_1(t) - b]^{+}$, for $T \geq 2$.}\newline

\noindent \textbf{Revenue comparison for the second price and the English auctions:} \cite{Milgrom&Weber82} shows that the English auction generates at least as much revenue as the second price auction in the symmetric interdependent values model with affiliated signals and risk neutral buyers. This ranking remains true for risk averse buyers as long as the utility function of each buyer has the form of a constant absolute risk aversion (\textit{CARA}) function; i.e., $u(x)=A(1-exp(-cx))$, where $A > 0$ and $c >0$. We now show that this ranking holds for the PLSC but not necessarily for the POSC.

For a given share fraction $\alpha$, a buyer values the resource in the PLSC as $\widetilde{u}(x) \triangleq u((1-\alpha )x)$; the function $\widetilde{u}(x)$ then determines his bids in both the English and the second price auction. If $u(x)$ is linear (as in the case of risk neutrality) or a CARA function, then $\widetilde{u}(x)$ is also linear or a CARA function, respectively. Let $t_{sp}(\alpha,y_1,z_1)$ denote the strategy function defined by \eqref{eq:t-plsc-sp} for the second price auction with the PLSC determined by $\alpha$ and $t_{eng}(\alpha,y_1,\mb{z})$ denote the strategy function defined by \eqref{eq:t-plsc-eng} for the English auction with the PLSC determined by $\alpha$. Assuming that $X_1, Y_1, Y_2, \ldots, Y_N$ are affiliated, the arguments in \cite{Milgrom&Weber82} implies that if $\widetilde{u}(x)$ is linear or a CARA function,
\beqn
t_{sp}(\alpha, y_1, z_1) \leq \E{t_{eng}(\alpha, y_1, z_1, \mb{Z}_{-1}) | Y_1 = y_1, Z_1 = z_1}.
\eeqn 
This inequality together with affiliation imply that for $y_1 > z_1$,
\begin{align*}
& t_{sp}(\alpha, z_1, z_1) \leq \E{t_{eng}(\alpha, z_1, z_1, \mb{Z}_{-1}) | Y_1 = z_1, Z_1 = z_1} \leq \E{t_{eng}(\alpha, z_1, z_1, \mb{Z}_{-1}) | Y_1 = y_1, Z_1 = z_1}, \\
& \Rightarrow \E{\alpha X_1 + (1-\alpha) t_{sp}(\alpha, Z_1, Z_1) | Y_1 > Z_1} \leq \E{\alpha X_1 + (1-\alpha) t_{eng}(\alpha, Z_1, \mb{Z}) | Y_1 > Z_1}, \\
& \Rightarrow R_{sp}^{plsc}(\alpha) \leq R_{eng}^{plsc}(\alpha).
\end{align*}
The expected total revenue from the English auction with a PLSC is therefore higher than the expected total revenue from the second price auction with the same PLSC.

In the POSC, however, a buyer uses the function $\widehat{u}(x) \triangleq u(x- \alpha \pos{x})$ to determine his bid which is not necessarily linear or CARA even if $u(x)$ has one of these properties. As a consequence, the revenue ranking of the English auction and the second price auction depends upon the underlying distribution of the random variables $(X_1,\mb{Y})$.\newline

\noindent \textbf{Limits on the choice of the share fraction:} Both the POSC and the PLSC share the property that the seller can obtain an arbitrarily large fraction of the surplus generated by the resource by choosing the share fraction $\alpha$ that is sufficiently close to one. We noted in Section \ref{sec:introduction}, however, that the choice of $\alpha$ may be constrained by the need to provide incentives to the winning buyer for the proper development of the resource; Section \ref{sec:pa} provides a formal analysis of PSCs on these lines. Here, we enumerate a number of other reasons that may constrain a seller in his choice of the share fraction $\alpha$.

A seller-nation, for instance, may be constrained by law or philosophy concerning its proper role in private enterprise. This is related to the incentives to the winning buyer, for a common argument against state-run businesses concerns their long-run incentives for efficient operation. In the 3G spectrum auction in India, for instance, the spectrum usage shares retained by the government are in the range of 3-8\% (see \cite{India3g}). It is increasingly common for nations to retain significant ownership shares in natural resources such as oil and minerals, e.g., in the 2005 Libyan auction, production shares retained by the govermnment ranged from 61.1-89.2\% (\cite{Cramton09}). A buyer may also have an outside option that bounds below the amount of profit he must expect to earn in order to participate in the auction. Finally, we model the seller as risk neutral; if he is instead strictly risk averse, then his optimal choice of the share fraction $\alpha$ will share the risk of the venture between the seller and the winning buyer.

An alternative perspective on our work is to interpret the share fraction $\alpha$ as a tax rate. Limited taxes on corporate earnings are common, and so a converse question must be asked: Do existing taxes on a corporation's earnings increase the government's revenue despite the fact that it diminishes the value of obtaining a resource to a corporate bidder and thereby decreases its bid in the auction? Our work shows that the revenue from a variety of forms of corporate income taxes can outweigh the negative effect of taxes on bids for government resources.\footnote{Unlike corporate income taxes, a PSC can be designed for a specific resource sale. It can also be used in conjunction with existing corporate taxes. See \cite{Grim&Schmidt2000} for an analysis of the effect of income taxes upon government auction revenue in an independent private values model with risk neutral buyers.}

\section{General Profit Sharing Contracts} \label{sec:general-psf}
This section leverages the intuition gained from Section \ref{sec:second-price} to study the revenue consequences of a general class of PSCs. The discussion here is restricted to the second price auction; the results, however, easily extend to the English auction, in the same way that Section \ref{sec:english} extends the results of Section \ref{sec:second-price} to the English auction.

Consider a PSC $\phi : \R \rightarrow \R$. If the winning buyer makes a payment~$b$ at the end of the auction stage and the value of the resource is revealed to be $x$, then the payment he makes to the seller in the profit sharing stage is $\phi(x-b)$. The function $\phi$ is an \textit{admissible PSC} if it satisfies the following properties:
\begin{enumerate}[(i)]
\item
$\phi(w)$ is nondecreasing in $w$ and $w-\phi(w)$ is increasing in $w$. 
\item
$\phi(0) = 0$ and $\lim_{w \rightarrow \infty} \phi(w) = \infty$.
\item
$\phi$ is continuous on $\R$.
\end{enumerate}
Property (i) says that the payment made to the seller in the profit sharing stage is nondecreasing in the preliminary profit. Moreover, the total profit of the winning buyer is increasing in the preliminary profit. Property (ii) says that if the preliminary profit is zero then no payment is made to the seller, and a large payment is made to the seller if the preliminary profit is large. Notice that $\phi(w) = \alpha\pos{w}$ corresponds to the POSC determined by $\alpha$ and $\phi(w) = \alpha w$ corresponds to the PLSC determined by $\alpha$. Both POSCs and PLSCs are admissible PSCs for $\alpha \in (0,1)$.

Given an admissible PSC $\phi$, define a function $s: \mc{I}_{Y}^2 \mapsto \R$, similar to \eqref{eq:s-posc-sp} and~\eqref{eq:t-plsc-sp}, as follows:
\beq{s-gen-sp}
s(y_1, z_1; \phi) \triangleq \left\{b: \E{u(X_1 - b - \phi(X_1-b)) | Y_1 = y_1, Z_1 = z_1} = 0 \right\}.
\eeq
Since $X_1 - b - \phi(X_1-b)$ is decreasing and continuous in~$b$, there is a unique $b$ that makes the expectation in \eqref{eq:s-gen-sp} equal to zero. Hence $s(y_1, z_1; \phi)$ is well-defined. As in the proof of Lemma~\ref{lemma:s-posc-sp-properties}, it can be shown that $s(y_1, z_1; \phi)$ is increasing in $y_1$ and nondecreasing in $z_1$. Moreover, paralleling the proof of Lemma \ref{lemma:eq-s-posc-sp}, the strategy $\beta_n(y_n) \triangleq s(y_n, y_n; \phi)$ for all~$n$ can be shown to constitute a symmetric BNE of the second price auction with an admissible PSC $\phi$. The seller's expected total revenue from the auction stage and the profit sharing stage combined, denoted by $R_{sp}(\phi)$, is:
\beq{rev-eqn-gen-sp}
R_{sp}(\phi) \triangleq \E{s(Z_1, Z_1; \phi) + \phi(X_1-s(Z_1, Z_1; \phi)) | Y_1 > Z_1}.
\eeq
Denote the revenue from the second price auction with only a one-time payment (i.e., $\phi(w) = 0$ for all $w$) by $R_{sp}(0) = \E{s(Z_1, Z_1; 0) | Y_1 > Z_1}$.\footnote{$\phi(w) = 0$ for all $w$ is not an admissible PSC as it does not satisfy $\lim_{w \rightarrow \infty} \phi(w) = \infty$.} Also, notice that $R_{sp}(0) = R_{sp}^{posc}(0) = R_{sp}^{plsc}(0)$, where $R_{sp}^{posc}$ is given by \eqref{eq:rev-eqn-posc-sp} and $R_{sp}^{plsc}$ is given by \eqref{eq:rev-eqn-plsc-sp}.

The Proposition \ref{proposition:rev-gen-sp} below summarizes the revenue consequences of the second price auction with an admissible PSC. The proof is outlined in Appendix \ref{sec:proof-rev-gen-sp}.
\begin{proposition} \label{proposition:rev-gen-sp}
The following statements holds for the second price auction with an admissible PSC $\phi$ and weakly risk averse buyers:
\begin{enumerate}[(i)]
\item
$R_{sp}(\phi) \geq R_{sp}(0) > 0$.
\item
If there exists some $\alpha \in (0,1)$ such that $\phi(w+\delta) - \phi(w) \leq \alpha \delta$ for all $w$ and $\delta$, then $R_{sp}(\phi) \leq R^{plsc}_{sp}(\alpha)$, where $R^{plsc}_{sp}(\alpha)$ is given by~\eqref{eq:rev-eqn-plsc-sp}. 
\end{enumerate}
\end{proposition}
As pointed out in Section \ref{sec:introduction}, there are many factors that can go into choosing a PSC. However, Proposition \ref{proposition:rev-gen-sp} shows that for any admissible PSC, the seller's expected total revenue from the two stages is positive and is higher than the revenue from the second price auction with only a one-time payment. Moreover, if the payment rate to the seller in the profit sharing stage is bounded from above by some $\alpha \in (0,1)$ (e.g, if $\phi^{'}(w) \leq \alpha$ for all $w$), then the seller's expected total revenue is bounded above by the PLSC determined by $\alpha$. This is a sense in which restricting attention to PLSCs among admissible PSCs need not diminish the expected revenue of the seller.

\section{Principal-Agent Relationship and PSCs} \label{sec:pa}
This section explores how the necessity of providing incentives to the winning buyer for developing the resource affects the choice of the share fraction $\alpha$ for the PLSC. As noted in Section \ref{sec:introduction}, the seller may need the expertise of the winning buyer in developing the resource. The effort of the winning buyer may not be fully observable to the seller. The auction and subsequent sharing contract are then components in a principal-agent relationship that must address the incentives provided to the winning buyer. While a comprehensive analysis of auctions with general PSCs in a principal-agent relationship is beyond the scope of this paper, we focus here on analyzing the revenue consequences of the PLSC in a principal-agent relationship. We first show that a one-time payment (i.e., $\alpha = 0$) can never be revenue optimal in the principal-agent relationship: a PLSC with a small share fraction~$\alpha$ gives a higher expected total revenue to the seller than the auction with only a one-time payment. Next, we show through an example that if sufficiently large gains can be realized by the winning buyer's effort, then the revenue maximizing share fraction $\alpha$ is strictly less than one. We conclude with some discussion on extending the analysis to the case of POSC. Again, the presentation here is restricted to the second price auction; the results, however, easily extend to the English auction.

Following \cite{MaCfee&McMillan86}, assume that the winning buyer exerts an effort $e \in \mc{E} \subset \R$ to obtain an additive improvement in the realized value of the resource; i.e., if buyer~$n$ wins and exerts an effort $e$, then the value of the resource to him is a realization of the random variable $X_n + e$. We use the term \textit{nominal value} to denote a realization of $X_n$ and \textit{effort-generated value} to denote a realization of $X_n + e$. The seller can verify the effort-generated value but not the effort~$e$ itself. The randomness in the nominal value conceals the winning buyer's effort from the seller, thereby preventing the seller from enforcing the effort that he deems optimal.\footnote{We assume in this section that $X_n$ is a continuous random variable. If $X_n$ is discrete, then the seller might infer the effort $e$ from knowing the discrete values that $X_n$ can take and the realization of $X_n + e$.} Let $c(e)$ be the monetary cost to the winning buyer of providing the effort $e$. The cost function $c$ is the same for each buyer $n$ and is known to the seller. However, since the effort $e$ is unverifiable, the cost $c(e)$ cannot be verified by the seller and hence cannot be included in the PLSC; the winning buyer bears the full cost of exerting an effort. If the winning buyer makes a payment $b$ at the end of the auction stage, exerts an effort $e$, and the nominal value of the resource to him is $x$, then the payment he makes in the second stage in the PLSC with share fraction $\alpha$ is $\alpha(x + e - b)$ and his total profit is:
\beq{payoff-plsc-pa}
x + e - b - \alpha(x + e - b) - c(e) = (1-\alpha)(x-b) + (1-\alpha)e - c(e). 
\eeq
This implies that the winning buyer's optimum effort must maximize $(1-\alpha)e - c(e)$ and can therefore be chosen independently of his signal as well as any other information he acquires from participating in the auction.

We make the following regularity assumptions on the cost function and the information structure in addition to those made in Section \ref{sec:model}:

\begin{assumption} \label{assumption:pa-topology}
The set $\mc{I}_{X}$ of possible values of the random variable $X_{1}$, the set $\mc{I}_{Y}$ of possible values of the random variable $Y_{1}$, and the set $\mc{E}$ of possible choices of effort $e$ are closed and bounded subintervals of $\mathbb{R}$.
\end{assumption}

\begin{assumption} \label{assumption:pa-continuity}
For any continuous function $h:\R \mapsto \R$, $a_n$'s, and $b_n$'s such that $a_n, b_n \in \mc{I}_{Y}$ and $a_n \leq b_n$ for all $n$, the conditional expectation $\E{h(X_1) | Y_1 \in [a_1,b_1], Y_2 \in [a_2,b_2], \ldots, Y_N \in [a_N,b_N]}$ is continuous in the $a_n$'s and $b_n$'s whenever it exists.
\end{assumption}

\begin{assumption} \label{assumption:pa-cost}
The cost function $c$ is nonnegative, increasing, convex, and continuously differentiable. Furthermore, the marginal cost $c'(e)$ satisfies $\inf_{e \in \mc{E}}c'(e) < 1$ and $\sup_{e \in \mc{E}}c'(e) > 1$.
\end{assumption}
Assumption \ref{assumption:pa-topology} insures that an optimum effort exists and is finite. Assumptions \ref{assumption:pa-topology} and \ref{assumption:pa-continuity} allow the derivate to be moved inside the expectation in the proofs in Appendix \ref{sec:proof-rev-plsc-sp-pa}. Assumption \ref{assumption:pa-cost} is standard for cost functions; it also avoids the trivial scenario where the optimum effort is always an end point of~$\mc{E}$ and does not depend on the share fraction $\alpha$. 

Because of Assumptions \ref{assumption:pa-topology} and \ref{assumption:pa-cost}, there exists for each $\alpha \in [0,1)$ a unique effort $e(\alpha)$ that maximizes $(1-\alpha)e - c(e)$. The optimum effort $e(\alpha)$ is nonincreasing in the share fraction $\alpha$; a higher share fraction $\alpha$ reduces the incentive for the winning buyer to exert effort to improve the value realized from the resource. Let $\kappa(\alpha)$ denote the net gain to the winning buyer from exerting the optimum effort:
\beq{pa-gain}
\kappa(\alpha) \triangleq \max_{e \in \mc{E}}\big((1-\alpha)e - c(e)\big) = (1-\alpha)e(\alpha) - c(e(\alpha)).
\eeq
The total profit of the winning buyer under the optimum effort, given that he makes a payment~$b$ at the end of the auction stage and the nominal value of the resource is $x$, equals $(1-\alpha)(x-b) + \kappa(\alpha)$. 

As in Section \ref{sec:plsc-sp}, we (re)define the function $t: [0,1) \times \mc{I}_{Y}^2 \rightarrow \R$ that will be used to characterize the bidding strategies of the buyers:
\beq{t-plsc-sp-pa}
t(\alpha, y_1, z_1) \triangleq \left\{b: \E{u\left((1-\alpha)(X_1-b) + \kappa(\alpha) \right) | Y_1 = y_1, Z_1 = z_1} = 0 \right\}.
\eeq
The function $t(\alpha, y_1, z_1)$ is well-defined. It is continuous in $y_1$ and $z_1$ by Assumption \ref{assumption:pa-continuity}, increasing in $y_1$, and nondecreasing in $z_1$. Paralleling the proof of Lemma \ref{lemma:eq-t-plsc-sp}, the strategy $\beta_n(y_n) \triangleq t(\alpha, y_n, y_n)$ for all~$n$ can be shown to constitute a symmetric BNE of the second price auction with the PLSC given by~$\alpha$ and the above principal-agent relationship. The seller's expected revenue from the auction stage is $\E{t(\alpha, Z_1, Z_1) | Y_1 > Z_1}$, his expected revenue from the PLSC stage is $\E{\alpha(X_1 + e(\alpha) - t(\alpha, Z_1, Z_1))| Y_1 > Z_1}$, and his expected total revenue from both stages under the principal-agent relationship, denoted by $R^{plsc}_{sp-pa}(\alpha)$, is:
\beq{rev-eqn-plsc-sp-pa}
R^{plsc}_{sp-pa}(\alpha) \triangleq \E{\alpha X_1 + \alpha e(\alpha) + (1-\alpha) t(\alpha, Z_1, Z_1) | Y_1 > Z_1},
\eeq
where $t(\alpha, y_1, z_1)$ is given by \eqref{eq:t-plsc-sp-pa}.

Proposition \ref{proposition:rev-plsc-sp-pa} below shows that the expected total revenue $R^{plsc}_{sp-pa}(\alpha)$ is strictly increasing in $\alpha$ for $\alpha$ near zero. Even with the necessity of providing incentives for effort to the winning buyer, it is therefore always the case that a PLSC with a small positive share fraction $\alpha$ produces a greater expected total revenue for the seller than a second-price auction with a one-time payment. The proof is in Appendix \ref{sec:proof-rev-plsc-sp-pa}.

\begin{proposition} \label{proposition:rev-plsc-sp-pa}
The following inequality holds for the second price auction with a PLSC and weakly risk averse buyers in the presence of the principal-agent relationship:
\beqn
\od{R^{plsc}_{sp-pa}(\alpha)}{\alpha}\big|_{\alpha = 0} > 0.
\eeqn
\end{proposition}

The following example provides insight into the expected revenue $R^{plsc}_{sp-pa}(\alpha)$ as a function of the share fraction $\alpha$. 

\begin{example} \label{eg:pa}
Consider two risk neutral buyers. The random variables $Y_1$ and $Y_2$, denoting the signals of the buyers, are independent and uniformly distributed in $[0,1]$. Given the signals $(y_1,y_2)$, the random variables $X_1$ and $X_2$, denoting the nominal value of the resource to the buyers, are independent and uniformly distributed in $[0,y_1 + y_2]$. By definition, $Z_1 = Y_2$. The cost function $c(e)$ equals $\gamma e^2$ for some $\gamma > 0$. 

It is straightforward to verify that
\beq{pa-ex-vars}
e(\alpha) = \frac{1-\alpha}{2\gamma}, 
\quad \kappa(\alpha) = \frac{(1-\alpha)^2}{4\gamma}, 
\quad t(\alpha,y_1,z_1) = \frac{y_1 + z_1}{2} + \frac{1-\alpha}{4\gamma}.
\eeq
Thus, buyer $1$ bids $y_1 + (1-\alpha)/(4\gamma)$ if his signal is $y_1$ and buyer $2$ bids $y_2 + (1-\alpha)/(4\gamma)$ if his signal is $y_2$. The seller's expected revenue from the auction stage is:
\beqn
\E{\min(Y_1, Y_2)} + \frac{1-\alpha}{4\gamma} = \frac{1}{3} + \frac{1-\alpha}{4\gamma}.
\eeqn
The seller's expected revenue from the PLSC stage is:
\beqn
\E{\alpha\left(X_1 + e(\alpha) - Y_2 - \frac{1-\alpha}{4\gamma}\right) \bigg| Y_1 > Y_2} 
= 2\alpha \E{\left(X_1 - Y_2 + \frac{1-\alpha}{4\gamma}\right)\indicator{Y_1 > Y_2}}.
\eeqn
Since $\E{X_1 \indicator{Y_1 > Y_2} | Y_1, Y_2} = (Y_1 + Y_2)\indicator{Y_1 > Y_2}/2$, the above expression can be further simplified to
\beqn
\frac{\alpha(1-\alpha)}{4\gamma} + \alpha\E{(Y_1 - Y_2)\indicator{Y_1 > Y_2}} 
= \frac{\alpha(1-\alpha)}{4\gamma} + \alpha\int_0^1\int_{y_2}^1(y_1 - y_2)dy_1dy_2 
= \frac{\alpha(1-\alpha)}{4\gamma} + \frac{\alpha}{6}.
\eeqn
The seller's expected total revenue is:
\beq{eg-pa-eq1}
R^{plsc}_{sp-pa}(\alpha) = \frac{1}{3} + \frac{1-\alpha}{4\gamma} + \frac{\alpha(1-\alpha)}{4\gamma} + \frac{\alpha}{6} = \frac{1}{3} + \frac{1}{4\gamma} + \frac{\alpha}{6} - \frac{\alpha^2}{4\gamma}.
\eeq
The right hand side of \eqref{eq:eg-pa-eq1} is maximized at $\alpha = \gamma/3$. 
\end{example}

We make the following observations concerning this example: 
\begin{enumerate}[(i)]
\item
If the cost parameter $\gamma$ is sufficiently small (i.e., $\gamma < 3$), then the share fraction $\alpha = \gamma/3$ that maximizes the seller's expected total revenue is strictly less than one. From~\eqref{eq:pa-ex-vars}, the optimum effort $e(\alpha)$ increases as $\gamma$ decreases. A small value of $\gamma$ models the case in which output per monetary cost of effort is relatively high. This results in a large effort-generated value of the resource and in turn higher bids in the auction stage. The seller in this case prefers to provide a greater incentive to develop the resource by choosing a small share fraction~$\alpha$ and reaps the benefits through a higher auction stage revenue together with a larger expected effort-generated value.

\item
If the cost parameter $\gamma$ is large (i.e., $\gamma \geq 3$), then the seller's expected total revenue is increasing in the share fraction $\alpha$ for $\alpha \in [0,1)$. In this case, the gain from exerting effort is small and the dependence of the revenue on the share fraction $\alpha$ is on the lines of the results of Section \ref{sec:plsc-sp}.
 
\item
The expected total revenue is strictly increasing in $\alpha$ for $\alpha \in [0,\min(\gamma/3,1))$. As shown more generally in Proposition \ref{proposition:rev-plsc-sp-pa}, the share fraction $\alpha = 0$ is never revenue optimal for $\gamma > 0$. 
\end{enumerate}

A complication arises in analyzing the auction with a POSC in the presence of the principal-agent relationship. A key simplification in the case of PLSC is that the optimum effort of the winning buyer depends on the share fraction $\alpha$ and the cost function but not on the signals or values of the buyers. The winning buyer therefore does not need to draw any inference from the auction stage in order to select his effort. In contrast to \eqref{eq:payoff-plsc-pa}, the winning buyer's total profit in a POSC is:
\beq{payoff-posc-pa}
x+e-b-\alpha \pos{x+e-b}-c(e).
\end{equation}
Because the sign of $x+e-b$ depends on the nominal value $x$, the optimum effort $e$ may depend upon $x$ or the winning buyer's beliefs about $x$.

We consider two alternative stages in the POSC at which the winning buyer chooses his effort. It is perhaps most plausible to assume that he chooses his effort after winning the auction but before the realization of the nominal value. His choice of effort in this case depends upon his beliefs about the nominal value, which may be influenced by any inference that he can draw from the price that he pays in the auction stage. Assuming that all buyers use the same symmetric increasing bidding strategy, the winning buyer can infer the second highest signal from the price. His choice of effort therefore depends upon the equilibrium bidding strategy, and the equilibrium bidding strategy depends upon the expected profit that will be generated when effort
is chosen optimally. The bidding strategy and the subsequent choice of the optimum effort are in this way tightly coupled and must be derived simultaneously, which is difficult.

Alternatively, consider the case in which the winning buyer observes the realization of his nominal value before choosing his effort. While this case is admittedly restrictive as a model of a principal and an agent, it retains the essential features of the hidden action and consequent moral hazard that distinguishes this relationship; we explore this special case in the hope that its results may generalize to richer models. The nominal value observed by the winning buyer outweighs any information that can be inferred from bids and the optimum effort again becomes independent of the bidding strategy. Appendix \ref{sec:pa-posc} analyzes the POSC in this case using the informational setup of Example \ref{eg:pa}. Three observations can be made concerning this example:

\begin{enumerate}[(i)]
\item
Numerical computations demonstrate that a POSC with a suitably small share fraction $\alpha$ generates a higher expected revenue for the seller than a one-time payment. This example therefore suggests that Proposition \ref{proposition:rev-plsc-sp-pa} extends to the case of the POSC. The calculations also demonstrate that the revenue-maximizing share fraction $\alpha$ for the seller is strictly less than one for sufficiently small values of the cost parameter $\gamma$. 

\item
The difference between the winning buyer's total profit in the POSC and the PLSC for the same share fraction $\alpha$ and effort $e$ is $\min\{0,\alpha(x+e-b)\}$, which is the fraction of the loss that the seller absorbs in the PLSC if the winning buyer suffers a loss. The marginal profit from exerting effort can only be larger in the POSC than in the PLSC. As a consequence, the winning buyer's optimum effort can only be larger in the POSC. A POSC is in this sense more effective for the seller than a PLSC as a means of eliciting effort from the winning buyer.

\item
The larger optimum effort by the winning buyer in the POSC produces a larger effort-generated value for the resource in comparison to the PLSC. As a consequence, the revenue superiority of the PLSC over the POSC can be reversed when the auction is followed by a principal-agent relationship. The reversal of the revenue ranking is observed for small values of $\gamma$ and large values of $\alpha$. For large values of the cost parameter $\gamma$, the gain from exerting effort is small and the dependence of the revenue on the share fraction $\alpha$ is on the lines of the results of Proposition~\ref{proposition:rev-plsc-posc-sp}.
\end{enumerate}

Point (i) supports this paper's emphasis on the advantages for the seller of profit-sharing in comparison to a one-time payment. Point (ii) holds in general for the POSC and the PLSC with this principal-agent model. Point (iii) is a noteworthy point concerning the relative properties of the POSC and the PLSC that may prove practically relevant to the seller's design of a PSC in the presence of the principal-agent relationship. The generality of points (i) and (iii) beyond Example~\ref{eg:pa} and for a more richer principal-agent model remains to be explored.


\section{Conclusions} \label{sec:conclusions}
We have shown for the second price auction and the English auction that a seller can increase his revenue from selling a resource by retaining a share of the ex-post positive profit realized from the resource. The seller can gain even further by committing ex-ante to share both losses and positive profits with the winning buyer. Our analysis is conducted for a symmetric interdependent values model that includes the cases of independent private values and a pure common value as special cases, and we consider a range of different sharing rules with which the seller retains a portion of the realized value of the resource. The generality of the sharing rule is important both because extraneous constraints may bind the seller in its selection or because the sharing must provide the winning buyer with the proper incentives for developing and reporting upon the resource. While such incentives considerations do influence the choice of the ex-post sharing rule, sharing a small fraction of losses and positive profits with the winning buyer is always better for the seller than no ex-post sharing.

\appendix

\section{Proof of Proposition \ref{proposition:rev-posc-sp}} \label{sec:proof-rev-posc-sp}
The first part of Proposition \ref{proposition:rev-posc-sp} follows trivially from Lemma \ref{lemma:s-posc-sp-properties}. Lemma \ref{lemma:ohlin} and Lemma \ref{lemma:rev-posc-sp-lemma1} provide the key steps for the second part.

\begin{lemma}[from \cite{Ohlin69}] \label{lemma:ohlin}
Let $W$ be a real random variable taking values in some interval $J_1$, and let $g_i : J_1 \mapsto J_2$ for $i = 1, 2$ be nondecreasing functions with values in some interval
$J_2$. Suppose $-\infty < \E{g_1(W)} = \E{g_2(W)} < \infty$. Let $h$ be a concave function such that $\E{h(g_i(W))}$ is well-defined for $i = 1, 2$. If there exists $w_0$ such that $g_1(w) \geq g_2(w)$ for $w < w_0$ and $g_1(w) \leq g_2(w)$ for $w > w_0$, then $\E{h(g_1(W))} \geq \E{h(g_2(W))}$. If $h$ is strictly concave and $\Prob{g_1(W) \neq g_2(W) } > 0$, then the inequality is strict.
\end{lemma}

\begin{lemma} \label{lemma:rev-posc-sp-lemma1}
For any given $y_1$ and $z_1$, $s(\alpha,y_1,z_1) + \alpha \CE{\pos{X_1-s(\alpha,y_1,z_1)}}{Y_1 = y_1, Z_1 = z_1}$ is nondecreasing in $\alpha$, where $s(\alpha,y_1,z_1)$ is defined by \eqref{eq:s-posc-sp}.
\end{lemma}
\begin{proof}
Consider $\wh{\alpha} > \alpha$, $\wh{\alpha} \in (0,1)$. To simplify the notation, define a random variable~$V$ whose probability distribution is identical to the conditional distribution of $X_1$ given $Y_1 = y_1$ and $Z_1 = z_1$. Let $\wh{b} \triangleq s(\wh{\alpha},y_1,z_1)$.

Since $b + \alpha \posb{V-b}$ is increasing in $b$, $b + \alpha \E{\posb{V-b}}$ is in turn increasing in $b$. There is a unique $\wt{b}$ that satisfies:
\beq{rev-posc-sp-lemma1-eq1a}
\wt{b} + \alpha \Ebig{\posb{V-\wt{b}}} = \wh{b} + \wh{\alpha} \Ebig{\posb{V-\wh{b}}}.
\eeq
To prove Lemma \ref{lemma:rev-posc-sp-lemma1}, it suffices to show that $s(\alpha,y_1,z_1) \leq \wt{b}$. From \eqref{eq:s-posc-sp}, this holds if and only if
\beq{rev-posc-sp-lemma1-eq1b}
\Ebig{u(V-\wt{b} - \alpha \posb{V-\wt{b}})} \leq 0.
\eeq

If $\wt{b} < \wh{b}$, then
\beqn
\wt{b} + \alpha \Ebig{\posb{V-\wt{b}}} 
< \wh{b} + \alpha \Ebig{\posb{V-\wh{b}}} 
\leq \wh{b} + \wh{\alpha} \Ebig{\posb{V-\wh{b}}},
\eeqn
which contradicts \eqref{eq:rev-posc-sp-lemma1-eq1a}. Hence $\wt{b} \geq \wh{b}$.

From \eqref{eq:rev-posc-sp-lemma1-eq1a},
\beqn
\Ebig{V - \wt{b} - \alpha \posb{V - \wt{b}}} = \Ebig{V - \wh{b} - \wh{\alpha} \posb{V-\wh{b}}}.
\eeqn
Moreover, 
\beqn
V - \wt{b} - \alpha \posb{V - \wt{b}} - (V - \wh{b} - \wh{\alpha} \posb{V-\wh{b}}) 
= \wh{b} - \wt{b} + \wh{\alpha} \posb{V-\wh{b}} - \alpha \posb{V - \wt{b}}.
\eeqn
Since $\wh{\alpha} > \alpha$ and $\wt{b} \geq \wh{b}$, the right hand side of the above equation is nonnegative for large positive values of~$V$, nonpositive for large negative values of $V$, and changes sign at most once. Lemma \ref{lemma:ohlin} then implies 
\beqn
\Ebig{u(V-\wt{b} - \alpha \posb{V-\wt{b}})} \leq \Ebig{u(V - \wh{b} - \wh{\alpha} \posb{V - \wh{b}})} = 0,
\eeqn
where the last equality follows from the definition of $\wh{b}$. This establishes \eqref{eq:rev-posc-sp-lemma1-eq1b} and the proof is complete.
\end{proof}

We can now prove part (ii) of Proposition \ref{proposition:rev-posc-sp}. From Lemma \ref{lemma:rev-posc-sp-lemma1}, for any $y_1 > z_1$ and $0 \leq \alpha < \wh{\alpha} <1$,
\begin{align} \label{eq:rev-posc-sp-lemma1-eq2}
s(\alpha,z_1,z_1) - s(\wh{\alpha},z_1,z_1)
& \leq \E{\wh{\alpha} \pos{X_1-s(\wh{\alpha},z_1,z_1)} - \alpha \pos{X_1-s(\alpha,z_1,z_1)} | Y_1 = z_1, Z_1 = z_1}, \nonumber \\
& \leq \E{\wh{\alpha} \pos{X_1-s(\wh{\alpha},z_1,z_1)} - \alpha \pos{X_1-s(\alpha,z_1,z_1)} | Y_1 = y_1, Z_1 = z_1}.
\end{align}
The expectation in the second line is conditioned on $Y_{1}=y_{1}$ and not $Y_{1}=z_{1}$. Lemma~\ref{lemma:s-posc-sp-properties} implies $s(\wh{\alpha},z_{1},z_{1})<s(\alpha,z_{1},z_{1})$ for $\wh{\alpha}>\alpha$, from which it follows that $\wh{\alpha}\pos{X_{1}-s(\wh{\alpha},z_{1},z_{1})} -\alpha\pos{X_{1}-s(\alpha,z_{1},z_{1})}$ is nondecreasing in $X_{1}$. The inequality \eqref{eq:rev-posc-sp-lemma1-eq2} then applies Lemma~\ref{lemma:fosd-sp}. Now,
\begin{align*}
R^{posc}_{sp}(\wh{\alpha}) & = \E{s(\wh{\alpha}, Z_1, Z_1) + \wh{\alpha}\pos{X_1-s(\wh{\alpha}, Z_1, Z_1)} | Y_1 > Z_1} \\
& = \E{\E{ s(\wh{\alpha}, Z_1, Z_1) + \wh{\alpha}\pos{X_1-s(\wh{\alpha}, Z_1, Z_1)} | Y_1, Z_1} | Y_1 > Z_1}, \\
& \geq \E{ \E{ s(\alpha, Z_1, Z_1) + \alpha\pos{X_1-s(\alpha, Z_1, Z_1)} | Y_1, Z_1} | Y_1 > Z_1}, \\
& = \E{ s(\alpha, Z_1, Z_1) + \alpha\pos{X_1-s(\alpha, Z_1, Z_1)} |  Y_1 > Z_1}, \\ 
& = R^{posc}_{sp}(\alpha),
\end{align*}
where the inequality is from \eqref{eq:rev-posc-sp-lemma1-eq2}. Finally, from Lemma~\ref{lemma:s-posc-sp-properties}, $s(0, z_1, z_1) > 0$. Hence, $R^{posc}_{sp}(0) > 0$ and $R^{posc}_{sp}(\alpha) \geq R^{posc}_{sp}(0) > 0$. This completes the proof. \eop

\section{Proof of Proposition \ref{proposition:rev-plsc-sp}} \label{sec:proof-rev-plsc-sp}
Part (i) follows trivially from Lemma \ref{lemma:t-plsc-sp-properties}. Turning to part (ii), the seller's expected revenue from the PLSC stage is:
\begin{align*}
\E{\alpha(X_1-t(\alpha, Z_1, Z_1))| Y_1 > Z_1} 
& = \E{\E{\alpha(X_1-t(\alpha, Z_1, Z_1))| Y_1, Z_1}| Y_1 > Z_1}, \\ 
& = \E{\alpha(\E{X_1 | Y_1, Z_1} - t(\alpha, Z_1, Z_1))| Y_1 > Z_1}. 
\end{align*}
For $Y_1 > Z_1$, Lemma \ref{lemma:t-plsc-sp-properties} implies that $\E{X_1 | Y_1, Z_1} \geq t(\alpha, Y_1, Z_1) > t(\alpha, Z_1, Z_1)$. The expected revenue from this tage is therefore positive.

The seller's expected total revenue is given by:
\begin{align} \label{eq:prop-rev-plsc-sp-eq1}
R^{plsc}_{sp}(\alpha) 
& = \E{t(\alpha, Z_1, Z_1) + \alpha(X_1-t(\alpha, Z_1, Z_1))| Y_1 > Z_1}, \nonumber \\ 
& = \E{ \E{\alpha X_1 + (1-\alpha) t(\alpha, Z_1, Z_1) | Y_1, Z_1} | Y_1 > Z_1}, \nonumber \\
& = \E{ \alpha \E{X_1 | Y_1, Z_1} + (1-\alpha) t(\alpha, Z_1, Z_1) | Y_1 > Z_1}.
\end{align}
Consider $0 \leq \alpha < \wh{\alpha} < 1$. Since $\E{X_1 | Y_1, Z_1} > t(\alpha, Z_1, Z_1)$ for all $\alpha \in [0,1)$ and $t(\alpha, Z_1, Z_1)$ is nondecreasing in $\alpha$, we get
\begin{align*}
\alpha \E{X_1 | Y_1, Z_1} + (1-\alpha) t(\alpha, Z_1, Z_1) 
& < \wh{\alpha} \E{X_1 | Y_1, Z_1} + (1-\wh{\alpha}) t(\alpha, Z_1, Z_1), \\ 
& \leq \wh{\alpha} \E{X_1 | Y_1, Z_1} + (1-\wh{\alpha}) t(\wh{\alpha}, Z_1, Z_1).
\end{align*}
Hence, $\alpha \E{X_1 | Y_1, Z_1} + (1-\alpha) t(\alpha, Z_1, Z_1)$ is increasing in $\alpha$. This along with \eqref{eq:prop-rev-plsc-sp-eq1} imply that the seller's total revenue $R^{plsc}_{sp}(\alpha)$ is increasing in $\alpha$. This completes the proof. \eop

\section{Proof of Proposition \ref{proposition:rev-plsc-posc-sp}} \label{sec:proof-rev-plsc-posc-sp}
We start with the following lemma:
\begin{lemma} \label{lemma:rev-plsc-posc-sp-lemma1}
For any $\alpha \in [0,1)$, $y_1$, and $z_1$,
\begin{multline} \label{eq:rev-plsc-posc-sp-lemma1-eq1}
\alpha \E{ X_1 | Y_1 = y_1, Z_1 = z_1} + (1-\alpha) t(\alpha, y_1, z_1) \\
\geq s(\alpha, y_1, z_1) + \alpha \E{\pos{X_1- s(\alpha, y_1, z_1)} | Y_1 = y_1, Z_1 = z_1},
\end{multline}
where $s(\alpha,y_1,z_1)$ is defined by \eqref{eq:s-posc-sp} and $t(\alpha,y_1,z_1)$ is defined by \eqref{eq:t-plsc-sp}.
\end{lemma}
\begin{proof}
From \eqref{eq:t-plsc-sp}, $\E{u((1-\alpha)(X_1-b))|Y_1 = y_1, Z_1 = z_1} = 0$ for $b = t(\alpha, y_1, z_1)$. Then \eqref{eq:rev-plsc-posc-sp-lemma1-eq1} is true if and only if
\beq{rev-plsc-posc-sp-lemma1-eq1a}
\begin{array}{l}
\E{u((1-\alpha)(X_1-\hat{b}))|Y_1 = y_1, Z_1 = z_1} \geq 0, \\ 
\displaystyle \text{for } \hat{b} = \frac{1}{(1-\alpha)}\left(s(\alpha, y_1, z_1) + \alpha \E{\pos{X_1- s(\alpha, y_1, z_1)} - X_1 | Y_1 = y_1, Z_1 = z_1}\right).
\end{array}
\eeq
Inequality \eqref{eq:rev-plsc-posc-sp-lemma1-eq1a} implies that $t(\alpha,y_{1},z_{1})\geq
\wh{b}$, which is equivalent to \eqref{eq:rev-plsc-posc-sp-lemma1-eq1}. To simplify the notation, define a random variable $V$ whose probability distribution is identical to the conditional distribution of $X_1 - s(\alpha, y_1, z_1)$ given $Y_1 = y_1$ and $Z_1 = z_1$. With this new notation, inequality \eqref{eq:rev-plsc-posc-sp-lemma1-eq1a} reduces to
\beq{rev-plsc-posc-sp-lemma1-eq1b}
\E{u\left((1-\alpha)V - \alpha \E{\pos{V} - V}\right)} \geq 0.
\eeq

The definition of $s(\alpha, y_1, z_1)$ implies that $\E{u(V-\alpha \pos{V})} = 0$. Also, 
\beqn
\E{V-\alpha \pos{V}} = \E{(1-\alpha)V - \alpha \E{\pos{V} - V}}.
\eeqn 
Since $\E{\pos{V} - V} \geq 0$, $\left(V-\alpha \pos{V}\right) - \left((1-\alpha)V - \alpha \E{\pos{V} - V}\right)$ is nonnegative for positive values of $V$, nonpositive for large negative values of $V$, and changes sign at most once. Lemma~\ref{lemma:ohlin} then implies
\beqn
\E{u\left((1-\alpha)V - \alpha \E{\pos{V} - V}\right)} \geq \E{u(V-\alpha \pos{V})} = 0.
\eeqn
This establishes \eqref{eq:rev-plsc-posc-sp-lemma1-eq1b} and the proof is complete.
\end{proof}

Next, we use Lemma \ref{lemma:rev-plsc-posc-sp-lemma1} to prove Proposition \ref{proposition:rev-plsc-posc-sp}. Notice that
\begin{align*}
R^{plsc}_{sp}(\alpha)
& = \E{\alpha X_1 + (1-\alpha) t(\alpha, Z_1, Z_1) | Y_1 > Z_1}, \\ 
& = \E{\E{\alpha X_1 + (1-\alpha) t(\alpha, Z_1, Z_1) | Y_1, Z_1} | Y_1 > Z_1},
\end{align*}
and
\begin{align*}
R^{posc}_{sp}(\alpha)
& = \E{s(\alpha, Z_1, Z_1) + \alpha\pos{X_1-s(\alpha, Z_1, Z_1)} | Y_1 > Z_1},\\
& = \E{\E{s(\alpha, Z_1, Z_1) + \alpha\pos{X_1-s(\alpha, Z_1, Z_1)} | Y_1, Z_1} | Y_1 > Z_1}.
\end{align*}
It suffices to show that for any $y_1 > z_1$,
\begin{multline*}
\E{\alpha X_1 + (1-\alpha) t(\alpha, z_1, z_1) | Y_1 = y_1, Z_1 = z_1} \\ \geq \E{s(\alpha, z_1, z_1) + \alpha\pos{X_1-s(\alpha, z_1, z_1)} | Y_1 = y_1, Z_1 = z_1},
\end{multline*}
or, equivalently,
\beq{rev-plsc-posc-sp-eq1}
\alpha \E{ X_1 - \pos{X_1-s(\alpha, z_1, z_1)} | Y_1 = y_1, Z_1 = z_1} \geq s(\alpha, z_1, z_1) - (1-\alpha) t(\alpha, z_1, z_1).
\eeq

From Lemma \ref{lemma:rev-plsc-posc-sp-lemma1},
\begin{align*}
s(\alpha, z_1, z_1) - (1-\alpha) t(\alpha, z_1, z_1) 
& \leq \alpha \E{X_1 - \pos{X_1- s(\alpha, z_1, z_1)}| Y_1 = z_1, Z_1 = z_1}, \\
& \leq \alpha \E{X_1 - \pos{X_1- s(\alpha, z_1, z_1)}| Y_1 = y_1, Z_1 = z_1}.
\end{align*}
The expectation in the second line is conditioned on $Y_{1}=y_{1}$ not $Y_{1}=z_{1}$. The second inequality follows from Lemma \ref{lemma:fosd-sp}. It implies \eqref{eq:rev-plsc-posc-sp-eq1} and completes the proof. \eop

\section{An Outline of the Proof of Proposition \ref{proposition:rev-gen-sp}} \label{sec:proof-rev-gen-sp}
To prove the first part of the claim, we start with the following lemma:
\begin{lemma} \label{lemma:proof-rev-gen-sp-lemma1}
For any admissible PSC $\phi$,
\beq{proof-rev-gen-sp-lemma1-eq1}
s(y_1,z_1;\phi) + \E{\phi(X_1-s(y_1,z_1;\phi)) | Y_1 = y_1, Z_1 = z_1} \geq s(y_1, z_1; 0).
\eeq
\end{lemma}
\begin{proof}
The solution $b$ to the equation $\CE{u(X_{1}-b)}{Y_{1}=y_{1},Z_{1}=z_{1}} = 0$ defines $s(y_{1},z_{1};0)$. Since $\CE{u(X_{1}-b)}{Y_{1}=y_{1},Z_{1}=z_{1}}$ is strictly decreasing in $b$, \eqref{eq:proof-rev-gen-sp-lemma1-eq1} is equivalent to
\beq{proof-rev-gen-sp-lemma1-eq2}
\E{u(X_1-s(y_1,z_1;\phi) - \E{\phi(X_1-s(y_1,z_1;\phi)) | Y_1 = y_1, Z_1 = z_1}) | Y_1 = y_1, Z_1 = z_1} 
\leq 0.
\eeq
Let $V$ be a random variable with probability distribution identical to the conditional distribution of $X_1-s(y_1,z_1;\phi)$ given $Y_1 = y_1$ and $Z_1 = z_1$. Then to establish inequality \eqref{eq:proof-rev-gen-sp-lemma1-eq2}, we equivalently need to show that $\E{u(V-\E{\phi(V)})} \leq 0$. The definition of $s(y_{1},z_{1};\phi)$ in \eqref{eq:s-gen-sp} implies that $\E{u(V-\phi(V))} = 0$. Notice that $\phi(V) - \E{\phi(V)}$ is nonnegative for large positive values of $V$, nonpositive for large negative values of $V$, and changes sign at most once. The proof is then completed by an application of Lemma~\ref{lemma:ohlin}.
\end{proof}

Lemma \ref{lemma:proof-rev-gen-sp-lemma1} and the properties of an admissible PSC can then be used to establish the first part of the claim. The proof is similar to the use of Lemma \ref{lemma:rev-posc-sp-lemma1} to prove Proposition \ref{proposition:rev-posc-sp} in the special case of $\alpha = 0$ and $\wh{\alpha} \in (0,1)$. The following lemma is used to prove the second part of the claim:

\begin{lemma} \label{lemma:proof-rev-gen-sp-lemma2}
Let $\phi$ satisfy the condition of part (ii) of Proposition \ref{proposition:rev-gen-sp}. Then,
\begin{multline} \label{eq:proof-rev-gen-sp-lemma2-eq1}
(1-\alpha)t(\alpha, y_1, z_1) + \alpha\E{X_1 | Y_1 = y_1, Z_1 = z_1} \geq \\ s(y_1,z_1;\phi) + \E{\phi(X_1-s(y_1,z_1;\phi)) | Y_1 = y_1, Z_1 = z_1},
\end{multline}
where $t(\alpha, y_1, z_1)$ is defined by \eqref{eq:t-plsc-sp}.
\end{lemma}
\begin{proof}
By the definition of $t(\alpha, y_1, z_1)$, inequality \eqref{eq:proof-rev-gen-sp-lemma2-eq1} is equivalent to
\begin{multline} \label{eq:proof-rev-gen-sp-lemma2-eq2}
\E{u((1-\alpha)X_1 - s(y_1,z_1;\phi) - \E{\phi(X_1-s(y_1,z_1;\phi)) - \alpha X_1 | Y_1 = y_1, Z_1 = z_1}) | Y_1 = y_1, Z_1 = z_1} 
\\ \geq 0.
\end{multline}
Let $V$ be the random variable defined as in the proof of Lemma \ref{lemma:proof-rev-gen-sp-lemma1}. Then to establish \eqref{eq:proof-rev-gen-sp-lemma2-eq2}, we equivalently need to show that $\E{u((1-\alpha)V-\E{\phi(V)-\alpha V})} \geq 0$, given $\E{u(V-\phi(V))} = 0$. Notice that $\phi(V) \leq \alpha V$ for $V \geq 0$ and $\phi(V) \geq \alpha V$ for $V \leq 0$. The proof is then completed by an application of Lemma~\ref{lemma:ohlin}, as in the proof of Lemma~\ref{lemma:rev-plsc-posc-sp-lemma1}. 
\end{proof}

Notice that $\alpha X_1 - \phi(X_1 - s(y_1,z_1;\phi))$ is nondecreasing in $X_1$. The second part of the claim is established using Lemma \ref{lemma:proof-rev-gen-sp-lemma2}, paralleling the use of Lemma \ref{lemma:rev-plsc-posc-sp-lemma1} to prove Proposition~\ref{proposition:rev-plsc-posc-sp}. \eop

\section{Proof of Proposition \ref{proposition:rev-plsc-sp-pa}} \label{sec:proof-rev-plsc-sp-pa}
The proof is by explicitly computing the derivative of $R^{plsc}_{sp-pa}(\alpha)$ at $\alpha = 0$. We first verify that: the derivate with respect to $\alpha$ can be taken inside the expectation in~\eqref{eq:rev-eqn-plsc-sp-pa}; and given any $y_1$ and $z_1$, the function $t(\alpha,y_1,z_1)$, define by \eqref{eq:t-plsc-sp-pa}, is differentiable in $\alpha$. 

Throughout the rest of this section, we call a function differentiable on a closed interval if it is differentiable in the interior of the interval, the right derivative exists at the left end point of the interval, and the left derivative exists at the right end point of the interval. By Assumption~\ref{assumption:pa-cost}, there exists $\epsilon > 0$ such that $\inf_{e \in \mc{E}}c'(e) < 1 - \epsilon$. In rest of this section, this $\epsilon$ is kept fixed and $\alpha \in [0,\epsilon]$ below.

\begin{lemma} \label{lemma:proof-pa-lemma1}
Given any $y_1$ and $z_1$, the function $t(\alpha,y_1,z_1)$, defined by \eqref{eq:t-plsc-sp-pa}, is continuously differentiable in $\alpha$. Moreover,
\beq{proof-pa-eq1}
\pd{t(\alpha, y_1, z_1)}{\alpha} = \left(\frac{b - e(\alpha)}{1-\alpha} - \frac{\E{X_1 u'\left((1-\alpha)(X_1 - b) + \kappa(\alpha) \right)|Y_1 = y_1 , Z_1 = z_1}}{(1-\alpha)\E{u'\left((1-\alpha)(X_1 - b) + \kappa(\alpha) \right)|Y_1 = y_1 , Z_1 = z_1}}\right)_{b = t(\alpha, y_1, z_1)}.
\eeq
\end{lemma}
\begin{proof}
For notational convenience, define a random variable $V$ whose probability distribution is identical to the conditional distribution of $X_1$ given $Y_1 = y_1$ and $Z_1 = z_1$. Define a function $\mu(\alpha,b)$ as follows:
\beq{proof-pa-eq2}
\mu(\alpha,b) \triangleq \E{u\big((1-\alpha)(V - b) + \kappa(\alpha)\big)}.
\eeq 
We first show that the $\mu(\alpha,b)$ is continuously differentiable and then use the implicit function theorem to establish that the $t(\alpha, y_1, z_1)$ is continuously differentiable with respect to $\alpha$. 

For $\alpha \in  [0,\epsilon]$ the optimum effort in an interior point of \mc{E} and $e(\alpha) = (c')^{-1}(1-\alpha)$. Hence $e(\alpha)$ and $\kappa(\alpha)$ are continuously differentiable in $\alpha$, as is $u\big((1-\alpha)(V - b) + \kappa(\alpha)\big)$. Given any $b$, define $M_1$ as:
\beq{proof-pa-eq3}
M_1 \triangleq \max_{\alpha \in [0, \epsilon], v \in \mc{I}_{X}} \left| \pd{u\big((1-\alpha)(v - b) + \kappa(\alpha)\big)}{\alpha}\right|.
\eeq
$M_1$ is well-defined and finite as it is the maximum of a continuous function over a compact set. From \eqref{eq:proof-pa-eq2}, given any $b$ and $0 \leq \wt{\alpha} < \wh{\alpha} \leq \epsilon$, 
\begin{align} \label{eq:proof-pa-eq4}
\left|\frac{\mu(\wh{\alpha},b) - \mu(\wt{\alpha},b)}{\wh{\alpha} - \wt{\alpha}}\right| 
& \leq \E{\left|\frac{u\big((1-\wh{\alpha})(V - b) + \kappa(\wh{\alpha})\big) - u\big((1-\wt{\alpha})(V - b) + \kappa(\wt{\alpha})\big)}{\wh{\alpha} - \wt{\alpha}}\right|}.
\end{align}
By the mean value theorem, for each $V$ there is an $\overline{\alpha} \in (\wt{\alpha},\wh{\alpha})$ such that 
\beq{proof-pa-eq5}
\frac{u\big((1-\wh{\alpha})(V - b) + \kappa(\wh{\alpha})\big) - u\big((1-\wt{\alpha})(V - b) + \kappa(\wt{\alpha})\big)}{\wh{\alpha} - \wt{\alpha}} = \pd{u\big((1-\alpha)(V - b) + \kappa(\alpha)\big)}{\alpha}\bigg|_{\alpha = \overline{\alpha}}.
\eeq
Combining \eqref{eq:proof-pa-eq3}-\eqref{eq:proof-pa-eq5}, we get
\beqn
\left|\frac{\mu(\wh{\alpha},b) - \mu(\wt{\alpha},b)}{\wh{\alpha} - \wt{\alpha}}\right| \leq M_1.
\eeqn
Then, by the dominated convergence theorem and \eqref{eq:proof-pa-eq2},
\beq{proof-pa-eq7}
\pd{\mu(\alpha,b)}{\alpha}
= \E{\pd{u\big((1-\alpha)(V - b) + \kappa(\alpha)\big)}{\alpha}}
= \E{( b + \kappa'(\alpha) - V)u'\left((1-\alpha)(V - b) + \kappa(\alpha) \right)}.
\eeq
A similar analysis shows that the function $\mu(\alpha,b)$ is differentiable in $b$. Moreover,
\beq{proof-pa-eq8}
\pd{\mu(\alpha,b)}{b} 
= \E{\pd{u\big((1-\alpha)(V - b) + \kappa(\alpha)\big)}{b}}
= -(1-\alpha)\E{u'\left((1-\alpha)(V - b) + \kappa(\alpha) \right)}.
\eeq
The right hand sides of \eqref{eq:proof-pa-eq7} and \eqref{eq:proof-pa-eq8} imply that $\mu(\alpha,b)$ is continuously differentiable and $\partial\mu(\alpha,b) / \partial b$ $\neq 0$. By definition, $\mu(\alpha,b) = 0$ for $b = t(\alpha, y_1, z_1)$. Then, by the implicit function theorem, $t(\alpha, y_1, z_1)$ is continuously differentiable in $\alpha$, and
\begin{align} \label{eq:proof-pa-eq9}
\pd{t(\alpha, y_1, z_1)}{\alpha} & = -\left(\frac{\pd{\mu(\alpha,b)}{\alpha}}{\pd{\mu(\alpha,b)}{b}}\right)_{b = t(\alpha, y_1, z_1)}, \nonumber \\
& = \left(\frac{b + \kappa'(\alpha)}{1-\alpha} - \frac{\E{X_1 u'\left((1-\alpha)(X_1 - b) + \kappa(\alpha) \right)|Y_1 = y_1 , Z_1 = z_1}}{(1-\alpha)\E{u'\left((1-\alpha)(X_1 - b) + \kappa(\alpha) \right)|Y_1 = y_1 , Z_1 = z_1}}\right)_{b = t(\alpha, y_1, z_1)}.
\end{align}
Finally, by the envelope theorem, $\kappa'(\alpha) = -e(\alpha)$. Substituting this in \eqref{eq:proof-pa-eq9} completes the proof.
\end{proof}

\begin{lemma} \label{lemma:proof-pa-lemma2}
For any $y_1$, $z_1$, and $\alpha$,
\beqn
(1-\alpha)\pd{t(\alpha, y_1, z_1)}{\alpha} - t(\alpha, y_1, z_1) + e(\alpha) + \E{X_1 | Y_1 = y_1, Z_1 = z_1} \geq 0,
\eeqn
where the $t(\alpha, y_1, z_1)$ is defined by \eqref{eq:t-plsc-sp-pa}.
\end{lemma}
\begin{proof}
From \cite{Esaryetal67}, a single random variable is associated; hence for any pair of nondecreasing functions $g_1$ and $g_2$ and any random variable $W$, $\E{g_1(W)g_2(W)} \geq \E{g_1(W)}\E{g_2(W)}$ whenever the expectations exist. For the random variable $X_1$, taking $g_1(X_1) = X_1$ and $g_2(X_1) = -u'\big((1-\alpha)(X_1 - b) + \kappa(\alpha) \big)$, and noticing that $-u'$ is an increasing function, we get
\beq{proof-pa-eq11}
\frac{\E{X_1 u'\left((1-\alpha)(X_1 - b) + \kappa(\alpha) \right)|Y_1 = y_1 , Z_1 = z_1}}{\E{u'\left((1-\alpha)(X_1 - b) + \kappa(\alpha) \right)|Y_1 = y_1 , Z_1 = z_1}} \leq \E{X_1|Y_1 = y_1 , Z_1 = z_1}.
\eeq
Combining Lemma \ref{lemma:proof-pa-lemma1} with \eqref{eq:proof-pa-eq11} completes the proof.
\end{proof}

Since $t(\alpha, y_1, z_1)$ is continuously differentiable in $\alpha$ and continuous in $y_1$ and $z_1$, $\partial t(\alpha, y_1, z_1)/ \partial \alpha$ is continuous in $\alpha$, $y_1$, and $z_1$. As noted in the proof of Lemma \ref{lemma:proof-pa-lemma1}, $e(\alpha)$ is continuously differentiable. Hence, the function $\lambda(\alpha, x_1, z_1) \triangleq \alpha x_1 + \alpha e(\alpha) + (1-\alpha) t(\alpha, z_1, z_1)$ is differentiable in $\alpha$ and the derivate is continuous in $\alpha$, $x_1$, and $z_1$. Using an analysis similar to the proof of Lemma \ref{lemma:proof-pa-lemma1} allows us to change the order of expectation and the derivative with respect to~$\alpha$ in the right hand side of \eqref{eq:t-plsc-sp-pa}, resulting in
\begin{align} \label{eq:proof-pa-eq12}
\pd{R^{plsc}_{sp-pa}(\alpha)}{\alpha}
& = \E{\pd{\big(\alpha X_1 + \alpha e(\alpha) + (1-\alpha) t(\alpha, Z_1, Z_1)\big)}{\alpha} \bigg| Y_1 > Z_1}, \nonumber \\
& = \E{ X_1 + e(\alpha) + \alpha e'(\alpha)  - t(\alpha, Z_1, Z_1) + (1-\alpha) \pd{t(\alpha, Z_1, Z_1)}{\alpha} \bigg| Y_1 > Z_1}.
\end{align}
To complete the proof of Proposition \ref{proposition:rev-plsc-sp-pa}, it suffices to show that for $y_1 > z_1$,
\beq{proof-pa-eq13}
\E{X_1|Y_1 = y_1, Z_1 = z_1} + e(0) - t(0, z_1, z_1) + \pd{t(\alpha, z_1, z_1)}{\alpha}\bigg|_{\alpha = 0} > 0.
\eeq
Using Lemma \ref{lemma:proof-pa-lemma2} with $\alpha = 0$ and $Y_1 = z_1$, we get
\begin{multline} \label{eq:proof-pa-eq14}
\E{X_1|Y_1 = y_1, Z_1 = z_1} + e(0) - t(0, z_1, z_1) + \pd{t(\alpha, z_1, z_1)}{\alpha}\bigg|_{\alpha = 0} > \\ \E{X_1|Y_1 = y_1, Z_1 = z_1} - \E{X_1|Y_1 = z_1, Z_1 = z_1}.
\end{multline}
Since $\E{X_1|Y_1 = y_1, Z_1 = z_1} > \E{X_1|Y_1 = z_1, Z_1 = z_1}$ for $y_1 > z_1$, \eqref{eq:proof-pa-eq13} holds and the proof is complete. \eop

\section{Example: Principal-Agent Relationship and POSC} \label{sec:pa-posc}
This section extends the analysis of Example \ref{eg:pa} to the case of a POSC. We assume that the winning buyer observes the nominal value of the resource before choosing his effort. Under this assumption, the equilibrium characterization of the second price auction with a POSC and the principal-agent relationship can be carried out for our general model with weakly risk averse buyers and a convex cost function. We restrict attention here to the setting of Example \ref{eg:pa} to numerically compute the revenue in the case of POSC and for simplicity of the presentation.

The winning buyer's total profit in the case of POSC is given by \eqref{eq:payoff-posc-pa}. Let $e(\alpha,x-b)$ denote his optimum effort, given the share fraction $\alpha$, the nominal value of the resource $x$, and the payment~$b$ in the auction stage. A case by case analysis can be used to solve for $e(\alpha,x-b)$,\footnote{It is helpful in this derivation to notice that $e-\gamma e^2$ is increasing for $e \leq 1/(2\gamma)$ and decreasing otherwise; and $(1-\alpha)e - \gamma e^2$ is increasing in $e$ for $e \leq (1-\alpha)/(2\gamma)$ and decreasing otherwise.} resulting in:
\beq{pa-posc-eq2}
e(\alpha,x-b) = \left\{ 
\begin{array}{l l}
\frac{1}{2\gamma} & \quad \text{if $x-b\leq -\frac{1}{2\gamma}$,}\\
b-x & \quad \text{if $x-b \in \left(-\frac{1}{2\gamma}, -\frac{1-\alpha}{2\gamma}\right)$,}\\
\frac{1-\alpha}{2\gamma} & \quad \text{if $x-b \geq -\frac{1-\alpha}{2\gamma}$.}
\end{array} \right.
\eeq
By contrast, the optimum effort in the case of PLSC is always equal to $(1-\alpha)/(2\gamma)$, which is smaller than the optimum effort $e(\alpha,x-b)$ in the case of POSC.

Let $\pi(\alpha, x-b)$ denote the corresponding maximum total profit of the winning buyer,
\begin{align}
\pi(\alpha, x-b) 
& \triangleq \max_{e\geq 0} [x + e - b - \alpha\pos{x + e - b} - \gamma e^2], \nonumber \\
& = x + e(\alpha,x-b) - b - \alpha\pos{x + e(\alpha,x-b) - b} - \gamma (e(\alpha,x-b))^2. \label{eq:pa-posc-eq1}
\end{align}
Since $x + e - b - \alpha\pos{x + e - b} - \gamma e^2$ is increasing in $x-b$, $\pi(\alpha, x-b)$ is increasing in $x-b$.

Paralleling \eqref{eq:t-plsc-sp-pa}, we (re)define the function $s: [0,1) \times \mc{I}_{Y}^2 \rightarrow \R$ that will be used to characterize the bidding strategies of the buyers in the case of POSC:
\beq{pa-posc-eq3}
s(\alpha, y_1, y_2) \triangleq \left\{b: \E{\pi(\alpha, X_1 - b) | Y_1 = y_1, Y_2 = y_2} = 0 \right\}.
\eeq
As in the proof of Lemma \ref{lemma:eq-s-posc-sp}, the strategy $\beta_n(y_n) \triangleq s(\alpha, y_n, y_n)$ for all~$n$ can be shown to constitute a symmetric BNE of the second price auction with a POSC given by~$\alpha$. The seller's expected total revenue from both stages, denoted by $R^{posc}_{sp-pa}(\alpha)$, is:
\beq{pa-posc-eq4}
R^{posc}_{sp-pa}(\alpha) \triangleq \E{s(\alpha, Y_2, Y_2) + \alpha\pos{X_1 + e\big(\alpha,X_1 - s(\alpha,Y_2,Y_2)\big) - s(\alpha,Y_2,Y_2)}| Y_1 > Y_2},
\eeq
where $s(\alpha, y_1, y_2)$ is given by \eqref{eq:pa-posc-eq3}.

\begin{figure}[t]
\begin{center}
\includegraphics[trim=0.5in 0.40in 0.5in 0.25in, clip=true, height=5.25in]{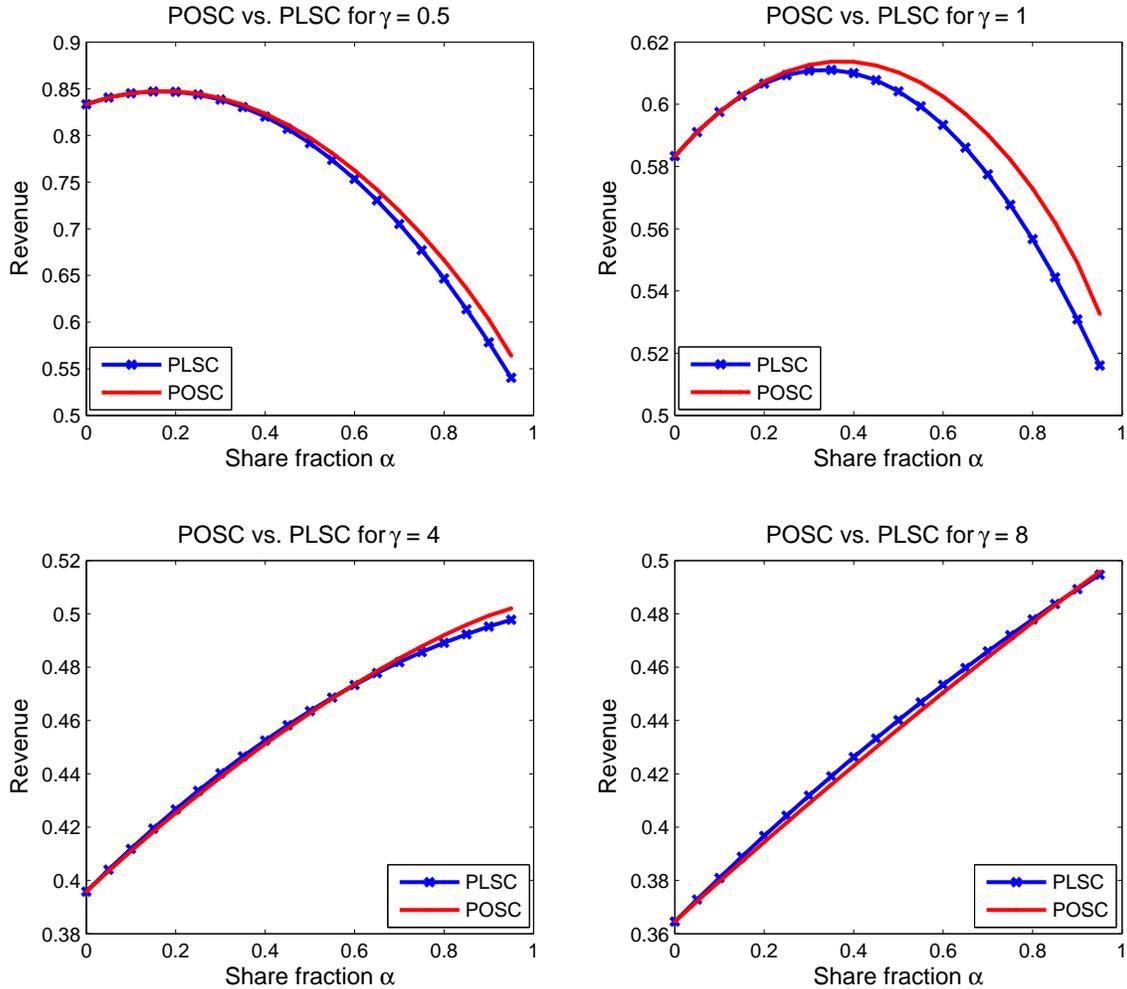}
\caption{\small \sl Revenue from the second price auction with a POSC and a PLSC as a function of the share fraction $\alpha$ under the principal-agent relationship. \label{fig:posc-plsc-pa-plot}} 
\end{center} 
\end{figure}

Using \eqref{eq:pa-posc-eq2} and \eqref{eq:pa-posc-eq3}, one can numerically compute $R^{posc}_{sp-pa}(\alpha)$. Figure \ref{fig:posc-plsc-pa-plot} plots the expected total revenue from the POSC and the PLSC for Example \ref{eg:pa} for different values of the cost parameter $\gamma$. Points (i)-(iii) at the end of the discussion of POSCs in Section \ref{sec:pa} follow directly from inspection of these figures. 

\bibliographystyle{abbrvnatold}
\bibliography{../../../../Citations/AuctionTheory,../../../../Citations/Misc}

\begin{thebibliography}{24}
\expandafter\ifx\csname natexlab\endcsname\relax\def\natexlab#1{#1}\fi

\bibitem[Abhishek et~al.(2011)Abhishek, Hajek, and
  Williams]{Abhishek-etal2011b}
V.~Abhishek, B.~Hajek, and S.~R. Williams.
\newblock On bidding with securities: Risk aversion and positive dependence.
\newblock Working paper. Manuscript at http://arxiv.org/abs/1111.1453, December
  2011.

\bibitem[Bikhchandani et~al.(2002)Bikhchandani, Haile, and
  Riley]{Bikhchandani2002}
S.~Bikhchandani, P.~A. Haile, and J.~G. Riley.
\newblock Symmetric separating equilibria in english auctions.
\newblock {\em Games and Economic Behavior}, 38\penalty0 (1):\penalty0 19 --
  27, 2002.

\bibitem[Cramton(1997)]{Cramton97}
P.~Cramton.
\newblock The {FCC} spectrum auctions: An early assessment.
\newblock {\em Journal of Economics and Management Strategy}, 6\penalty0
  (3):\penalty0 431--495, 1997.

\bibitem[Cramton(2009)]{Cramton09}
P.~Cramton.
\newblock How best to auction natural resources.
\newblock In P.~Daniel, B.~Goldsworthy, M.~Keen, and C.~McPherson, editors,
  {\em Handbook of Oil, Gas And Mineral Taxation}, chapter~10. International
  Monetary Fund, Washington, DC, 2009.

\bibitem[Cr{\'e}mer and McLean(1985)]{Cremer&McLean85}
J.~Cr{\'e}mer and R.~P. McLean.
\newblock Optimal selling strategies under uncertainty for a discriminating
  monopolist when demands are interdependent.
\newblock {\em Econometrica}, 53\penalty0 (2):\penalty0 345--361, 1985.

\bibitem[Cr{\'e}mer and McLean(1988)]{Cremer&McLean88}
J.~Cr{\'e}mer and R.~P. McLean.
\newblock Full extraction of the surplus in bayesian and dominant strategy
  auctions.
\newblock {\em Econometrica}, 56\penalty0 (6):\penalty0 1247--1257, 1988.

\bibitem[DeMarzo et~al.(2005)DeMarzo, Kremer, and Skrzypacz]{DeMarzo2005}
P.~M. DeMarzo, I.~Kremer, and A.~Skrzypacz.
\newblock Bidding with securities: Auctions and security design.
\newblock {\em American Economic Review}, 95\penalty0 (4):\penalty0 936--959,
  2005.

\bibitem[{Department of Telecommunications}(2008)]{India3g}
{Department of Telecommunications}.
\newblock Auction of {3G \& BWA} spectrum-information memorandum details.
\newblock Technical report, Government of India, December 2008.

\bibitem[Esary et~al.(1967)Esary, Proschan, and Walkup]{Esaryetal67}
J.~D. Esary, F.~Proschan, and D.~W. Walkup.
\newblock Association of random variables, with applications.
\newblock {\em The Annals of Mathematical Statistics}, 38\penalty0
  (5):\penalty0 1466--1474, 1967.

\bibitem[{Federal Communications Commission}(2007)]{FCCAuction73}
{Federal Communications Commission}.
\newblock Auction 73 for 700 {MHz} band.
\newblock Technical report, United States Government, October 2007.

\bibitem[Grimm and Schmidt(2000)]{Grim&Schmidt2000}
V.~Grimm and U.~Schmidt.
\newblock Revenue equivalence and income taxation.
\newblock {\em Journal of Economics and Finance}, 24:\penalty0 56--63, 2000.

\bibitem[Hansen(1985)]{Hansen85}
R.~G. Hansen.
\newblock Auctions with contingent payments.
\newblock {\em The American Economic Review}, 75\penalty0 (4):\penalty0
  862--865, 1985.

\bibitem[Krishna(2002)]{Krishna2002}
V.~Krishna.
\newblock {\em Auction Theory}.
\newblock Academic Press, March 2002.

\bibitem[Laffont and Tirole(1987)]{Laffont&Tirole87}
J.-J. Laffont and J.~Tirole.
\newblock Auctioning incentive contracts.
\newblock {\em The Journal of Political Economy}, 95\penalty0 (5):\penalty0
  921--937, 1987.

\bibitem[McAfee and McMillan(1986)]{MaCfee&McMillan86}
R.~P. McAfee and J.~McMillan.
\newblock Bidding for contracts: A principal-agent analysis.
\newblock {\em The RAND Journal of Economics}, 17\penalty0 (3):\penalty0
  326--338, 1986.

\bibitem[McAfee et~al.(1989)McAfee, McMillan, and Reny]{McAfee&McMillan&Reny89}
R.~P. McAfee, J.~McMillan, and P.~J. Reny.
\newblock Extracting the surplus in the common-value auction.
\newblock {\em Econometrica}, 57\penalty0 (6):\penalty0 1451--1459, 1989.

\bibitem[McAfee and Reny(1992)]{McAfee&Reny92}
R.~P. McAfee and P.~J. Reny.
\newblock Correlated information and mechanism design.
\newblock {\em Econometrica}, 60\penalty0 (2):\penalty0 395--421, 1992.

\bibitem[Mezzetti(2007)]{Mezzetti07}
C.~Mezzetti.
\newblock Mechanism design with interdependent valuations: Surplus extraction.
\newblock {\em Economic Theory}, 31:\penalty0 473--488, 2007.

\bibitem[Milgrom(2004)]{Milgrom04}
P.~Milgrom.
\newblock {\em Putting Auction Theory to Work}.
\newblock Cambridge University Press, 2004.

\bibitem[Milgrom and Weber(1982)]{Milgrom&Weber82}
P.~R. Milgrom and R.~J. Weber.
\newblock A theory of auctions and competitive bidding.
\newblock {\em Econometrica}, 50\penalty0 (5):\penalty0 1089--1122, 1982.

\bibitem[Ohlin(1969)]{Ohlin69}
J.~Ohlin.
\newblock On a class of measures of dispersion with application to optimal
  insurance.
\newblock {\em Astin Bulletin}, 5:\penalty0 249–--266, 1969.

\bibitem[Porter(1995)]{Porter95}
R.~H. Porter.
\newblock The role of information in {U.S.} offshore oil and gas lease auction.
\newblock {\em Econometrica}, 63\penalty0 (1):\penalty0 1--27, 1995.

\bibitem[Riley(1988)]{Riley88}
J.~G. Riley.
\newblock Ex post information in auctions.
\newblock {\em The Review of Economic Studies}, 55\penalty0 (3):\penalty0
  409--429, 1988.

\bibitem[Wilson(1987)]{Wilson87}
R.~Wilson.
\newblock Game-theoretic analyses of trading processes.
\newblock In T.~F. Bewley, editor, {\em Advances in Economic Theory: Fifth
  World Congress}, chapter~2. Cambridge University Press, Cambridge, 1987.

\end{thebibliography}

\end{document}